\documentclass[preprint,authoryear,12pt]{elsarticle}

\usepackage{epsfig}

\usepackage{amssymb}

\newcommand{\pd}{\partial}

\newcommand{\bdot}{\mbox{\boldmath $\cdot$}}

\newcommand{\curl}{\mbox{\boldmath $\nabla \times$}}

\newcommand{\cross}{\mbox{\boldmath $\times$}}

\newcommand{\surf}{\mbox{\boldmath ${\cal S}$}}
\newcommand{\bell}{\mbox{\boldmath ${\cal \ell}$}}

\newcommand{\vv}{{\bf v}}
\newcommand{\BB}{{\bf B}}

\newcommand{\uvr}{\mbox{\boldmath $\hat{r}$}}

\newcommand{\uvp}{\mbox{\boldmath $\hat{\phi}$}}

\journal{Advances in Space Research}

\begin{document}

\begin{frontmatter}



\title{A Three-Dimensional Babcock-Leighton Solar Dynamo Model: Initial Results with Axisymmetric Flows}


\author[label1]{Mark S. Miesch\corref{cor}}
\cortext[cor]{Corresponding author}
\ead{miesch@ucar.edu}


\author[label1,label2]{Kinfe Teweldebirhan}
\ead{kinntt2000@gmail.com}

\address[label1]{High Altitude Observatory, National Center for Atmospheric Research, 3080 Center Green Dr., Boulder, CO 80301}
\address[label2]{Physics Department, Faculty of Natural Sciences, Addis Ababa University, Box 1176, Addis Ababa, Ethiopia}

\begin{abstract}
The main objective of this paper is to introduce the STABLE (Surface flux Transport And Babcock-LEighton) solar dynamo model.  STABLE is a 3D Babcock-Leighton/Flux Transport dynamo model in which the source of poloidal field is the explicit emergence, distortion, and dispersal of bipolar magnetic regions (BMRs).  Here we describe the STABLE model in more detail than we have previously and we verify it by reproducing a 2D mean-field benchmark.  We also present some representative dynamo simulations, focusing on the special case of kinematic magnetic induction and axisymmetric flow fields.  Not all solutions are supercritical; it can be a challenge for the BL mechanism to sustain the dynamo when the turbulent diffusion near the surface is $\geq 10^{12}$ cm$^2$ s$^{-1}$.  However, if BMRs are sufficiently large, deep, and numerous, then sustained, cyclic, dynamo solutions can be found that exhibit solar-like features.  Furthermore, we find that the shearing of radial magnetic flux by the surface differential rotation can account for most of the net toroidal flux generation in each hemisphere, as has been recently argued for the Sun by \cite{camer15}.
\end{abstract}


\end{frontmatter}

\parindent=0.5 cm

\section{Introduction}
\label{sec:intro}

Over the last two decades, Babcock-Leighton (BL) dynamo models have emerged as a leading paradigm for explaining the origin of the solar activity cycle \citep{dikpa09,charb10,karak14}. The defining characteristic of BL models is the critical role of magnetic flux emergence and dispersal in the operation of the dynamo. Emerging flux structures appear in the solar photosphere as bipolar magnetic regions (BMRs) with systematic orientations such that the trailing flux (in the sense of rotational motion) is displaced poleward relative to the leading flux \citep{stenf12,mccli13}.   The subsequent evolution of this flux in response to differential rotation, meridional circulation, and convection, generates a mean poloidal field that can ultimately reverse the Sun's dipole moment, at least at the surface (see \S\ref{sec:flagship}).  The preferential tilt of BMRs is known as Joy's law and this process of poloidal field generation is known as the Babcock-Leighton (BL) mechanism.

Most current BL dynamo models may also be classified as Flux Transport dynamo (FTD) models in which the meridional circulation (MC) regulates the cycle period \citep{dikpa09,charb10,karak14}.  The equatorward migration of toroidal flux inferred from the solar butterfly diagram (\S\ref{sec:flagship}) is attributed to an equatorward flow of 2-3 m s$^{-1}$ near the base of the convection zone (CZ) where the progenitor flux for BMRs is thought to originate.   Though this presumed equatorward flow at the base of the CZ has not yet been detected, it has long been inferred based on the observed poleward circulation in the upper CZ and the constraint of mass conservation.  Recent photospheric observations and helioseismic inversions have called into question this simple single-celled picture of the MC \citep{hatha12b,zhao13,jacki15,rajag15}.  However, FTD models are still viable as long as the circulation at the base of the CZ is equatorward and convection contributes to the transport of poloidal flux across the CZ \citep{hazra14b,beluc15}.  Both conditions are supported by theory and global convection simulations, even when the overall MC profile is multi-cellular \citep{miesc05,miesc12b,passo15}.

The BL mechanism has a solid empirical grounding; we observe it operating in the solar photosphere as BMRs continually emerge and disperse.  The amount of flux emerging is more than enough to reverse the polar fields, at least at the surface, and there is evidence that the strength of the following cycle is correlated with the BL poloidal source term and with the strength of the Sun's polar fields during cycle minimum, as predicted \citep{schat78,svalg05,dasi10,munoz12}.  Other observed solar cycle features that are well reproduced by BL/FTD models include the equatorward migration of toroidal flux (solar butterfly diagram), the phase relationship between toroidal and poloidal fields, the phase coherence across grand minima, and the flux budget in active regions \citep{dikpa09,charb10,karak14,camer15}.

Though global MHD simulations of convective dynamos have made great strides in recent years \citep[e.g.][]{charb14}, they still cannot capture the full multi-scale complexity of flux emergence and the BL mechanism.  Until they do, hybrid approaches are necessary to model the solar cycle with maximum fidelity.

We describe one such hybrid approach here.  We call it the STABLE (Surface flux Transport And Babcock-LEighton) solar dynamo model.  STABLE is an FTD model but unlike most previous FTD models that only address the axisymmetric (2D) magnetic field components, STABLE is fully 3D.  So, it can capture the 11-year solar cycle as well as the explicit distortion and dispersal of photospheric BMRs that underlies the BL mechanism.  The latter capability in effect makes STABLE a 3D generalization of 2D (latitude/longitude) surface flux transport (SFT) models, which have had notable success in capturing the observed evolution of photospheric fields \citep{devor84,wang91b,schri03,bauma04,camer10,upton14a,jiang14a,jiang14b,hickm15}.  A 3D, kinematic FTD/SFT model similar to STABLE was recently described by \citet{yeate13}.  Meanwhile, \cite{lemer15} describe a somewhat different approach to unifying SFT models and FTD models based on retaining the 2D nature of each class of model and coupling them, so that each model provides the source term that sustains the other.  There have also been a few 3D mean-field dynamo models based on a turbulent $\alpha$-effect as opposed to the BL mechanism \citep[e.g.][]{chan04}.

One of the main advantages of the 3D approach over 2D FTD models is the potential for a more realistic depiction of flux emergence.  Though we acknowledge that this potential has not yet been fully realized, substantial progress in this direction has been made by \cite{yeate13} who employ a kinematic flow that lifts and twists toroidal magnetic fields, mimicking the effects of magnetic buoyancy.  In the future we will implement this and other flux emergence algorithms into STABLE.

Another promising reason to develop a 3D FTD model is the potential for data assimilation (DA).  If solar dynamo models are to be used for the prediction of future solar activity, they must assimilate observational data.  If the model is axisymmetric, this data assimilation typically makes use of the mean radial magnetic field at the surface of the sun as a function of latitude and time \cite[e.g.][]{dikpa07,jiang13}.  This is effectively 1D DA.  However, SFT models can exploit the observations more fully, assimilating the full observed radial magnetic field as a function of latitude, longitude and time (2D DA; see references above).  Thus, SFT models are able to model the time-evolving surface magnetic field of the Sun with more fidelity.  However, since they are not dynamo models, their predictive potential is relatively short-term, spanning less than a decade.  STABLE will be capable of assimilating complete 2D magnetograms (2D DA) for use with both short-term and long-term solar activity forecasting.  The 2D surface fields will also provide boundary conditions for corona and heliosphere models.

There are many other reasons for developing a 3D, nonlinear, MHD FTD model.  Another is turbulent transport.  The 3D formulation will allow us to replace turbulent diffusion and magnetic pumping with 3D convective flow fields either computed self-consistently or derived from observations.  It will also allow us to capture magneto-shear instabilities in the solar tachocline which may induce non-axisymmetric patterns in flux emergence and may contribute to poloidal field generation via an $\alpha$-effect \citep{gilma97,dikpa99,dikpa01c}.  Other sources of non-axisymmetric activity include converging flows into active regions \citep{camer12} and longitudinal variations in the meridional circulation.

First results from STABLE were reported by \cite{miesc14}, hereafter MD14.  In this paper we describe STABLE in somewhat more detail and consider the operation of the dynamo for the special case when the imposed flow fields are kinematic and axisymmetric.  In this case the mean (axisymmetric) induction equation decouples from the non-axisymmetric components and behaves essentially as a 2D FTD model.  This serves to verify the model and to provide a baseline for future simulations that will include non-axisymmetric flows and Lorentz-force feedbacks.  After describing the formulation of the model in \S\ref{sec:STABLE}, we test it against a 2D FTD benchmark in \S\ref{sec:verification}.  We then give an illustrative example of a solar dynamo simulation in \S\ref{sec:flagship} and we discuss the challenge of achieving self-sustained dynamo action in \S\ref{sec:supercritical}.  We close by summarizing our main results in \S\ref{sec:summary}.

\section{The STABLE Solar Dynamo Model}
\label{sec:STABLE}

\subsection{Kinematic Induction}\label{sec:kin}

The STABLE model solves the kinematic magnetohydrodynamic (MHD) induction equation in a 3D, rotating, spherical shell:
\begin{equation}\label{eq:indy}
\frac{\pd \BB}{\pd t} = \curl \left(\vv \cross \BB - 
\eta_t \curl \BB\right)
\end{equation}
where $\vv$ and $\BB$ are the velocity and magnetic field in the rotating reference frame and $\eta_t(r)$ is a turbulent diffusion.  This equation is solved by means of the Anelastic Spherical Harmonic (ASH) code, which currently serves as the {\em dynamical core} for the STABLE model.  ASH is a well established, parallel, MHD code that has been used extensively to study convection, instabilities, tachocline confinement, and other aspects of solar and stellar internal dynamics \citep{clune99,brun04,miesc05,miesc09,brun10b}.  For the initial stages of STABLE development, we have modified ASH to operate in a kinematic regime (the full MHD system will be considered in future work).

Though we will consider 3D and time-dependent flows in the near future, here we focus on steady, axisymmetric mean flows such that 
\begin{equation}\label{eq:flows}
\vv = \tilde{\rho}(r)^{-1} ~ \curl \left[\psi(r,\theta) \uvp\right] + \lambda \Omega(r,\theta) \uvp
\end{equation}
where $\psi(r,\theta)$ is the stream function for the meridional mass flux, $\tilde{\rho}(r)$ is the dimensionless density stratification,  $\lambda = r \sin\theta$ is the cylindrical radius, and $\Omega(r,\theta)$ is the differential rotation.  Note that there is no explicit $\alpha$-effect; the poloidal field generation needed to sustain the dynamo occurs as a consequence of the spot deposition algorithm described in section \ref{sec:spotmaker}.

\begin{figure}
\begin{center}
\includegraphics*[width=\linewidth]{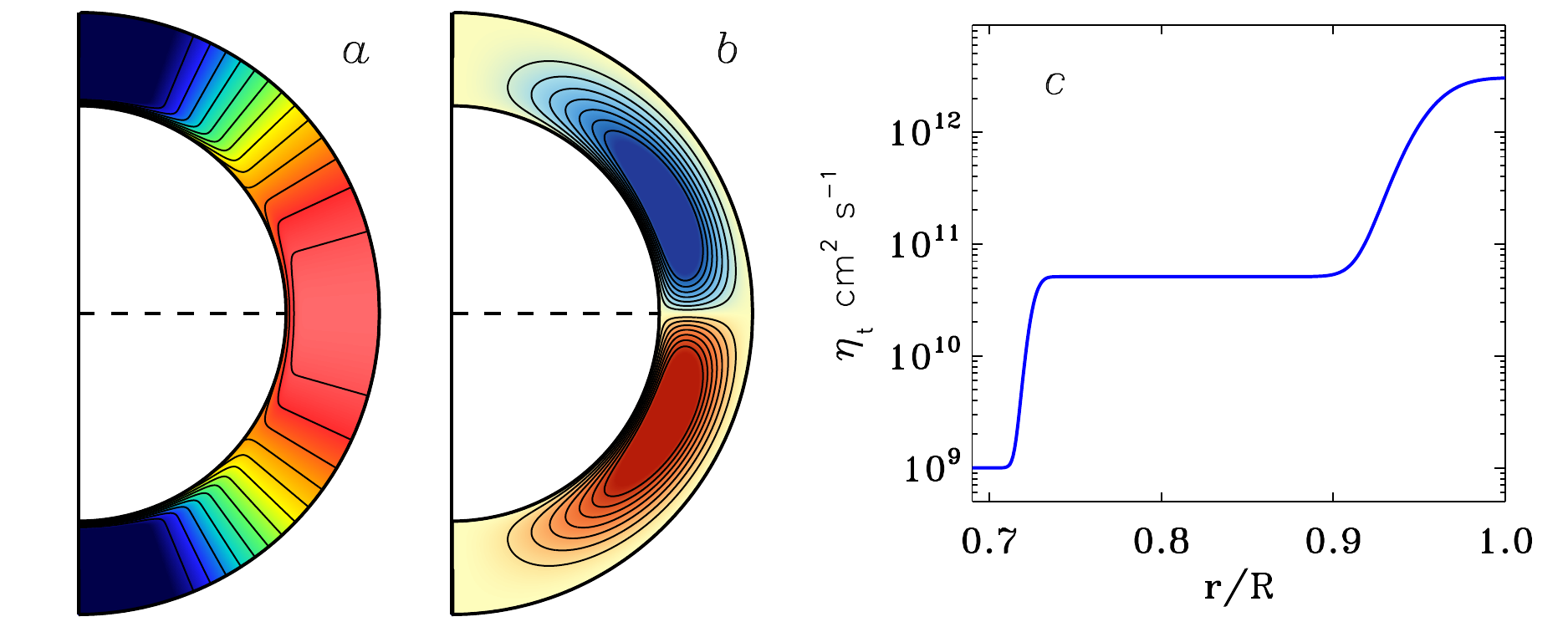}
\end{center}
\caption{Model components; Mean flows and turbulent diffusion. 
($a$) Angular velocity $\Omega/2\pi$, with color table
ranging from 350-480 nHz (blue to red/pink). ($b$) Meridional circulation, shown as 
streamlines of the mass flux, with red and blue denoting clockwise
and counter-clockwise circulation respectively.  Horizontal velocity 
amplitudes are approximately 14 m s$^{-1}$ and 1.5 m s$^{-1}$ in the
upper and lower CZ. ($c$) turbulent magnetic diffusivity $\eta_t$.
\label{fig:ingredients}}
\end{figure}

Note also that the use of an axisymmetric velocity field as given by eq.\ (\ref{eq:flows}) implies that the axisymmetric ($m=0$, where $m$ is the azimuthal wave number) component of equation (\ref{eq:indy}) decouples from the non-axisymmetric components ($m \neq 0$).  In other words, the mean (longitudinally-averaged) induction equation for the STABLE model is equivalent to a 2D FTD model when $\vv$ is axisymmetric and kinematic.  This allows us to make contact with previous 2D FTD models in the literature before moving on to more general 3D, nonlinear flow fields.

This point can be appreciated by averaging eq.\ (\ref{eq:indy}) over $\phi$, noting that $\vv$, $\eta_t$, and the curl operator are all independent of $\phi$.  Though the SpotMaker algorithm (\S\ref{sec:spotmaker}) is distinct from previous representations of the BL mechanism in the literature, one could in principle devise an axisymmetric source term that is equivalent from the perspective of the mean fields.  Thus, in terms of the evolution of the mean fields, the models presented here are similar to the 2D FTD models presented by \cite{nandy01} and \cite{munoz10}, who used an axisymmetric spot deposition algorithm.   However, unlike 2D FTD models, STABLE also provides the corresponding time evolution of the 2D (latitude/longitude) surface field.

In order to build on previous work, we specify $\vv(r,\theta)$ and $\eta_t(r)$ as in previous FTD models.  In particular, the $\Omega$ profile is taken from \citet{dikpa99b}
\begin{equation}\label{omega1}
\Omega(r,\theta) = \Omega_c + \frac{1}{2} \left[1 + \mbox{erf}\left(2 ~ \frac{r - r_c}{d}\right)\right]
\left(\Omega_s(\theta) - \Omega_c\right)
\end{equation}
where $\Omega_c = 2\pi \nu_c$ and
\begin{equation}\label{omega2}
\Omega_s(\theta) = 2\pi ~ (\nu_{eq} + a_2 \cos^2\theta + a_4 \cos^4\theta) ~~~.
\end{equation}
Here we use $\nu_c$ = 432.8 nHz, $\nu_{eq}$ = 460.7 nHz, $a_2 = $ - 62.9 nHz, $a_4$ = -67.13 nHz, $r_c = 0.7 R$, and $d = 0.05 R$, where $R$ is the solar radius.  The resulting profile is illustrated in Fig.\ \ref{fig:ingredients}$a$.

The meridional flow is the same as that used by \cite{dikpa10b}
and \cite{dikpa11}:
\begin{equation}\label{psim}
\psi(r,\theta) = - \psi_0 
~ \lambda^{-1} ~ \left(\theta - \theta_0\right) ~ f_{mc}(r) ~ h_{mc}(r,\theta)
\end{equation}
with
\begin{equation}\label{fmc}
f_{mc}(r) = \sin\left[\frac{\pi (\tilde{r} - r_b)}{(\tilde{R} - r_b)}\right] 
\exp\left[-\left(\frac{\tilde{r}-r_0}{\Gamma}\right)^2\right]
\end{equation}
and
\begin{equation}\label{hmc}
h_{mc}(r,\theta) = \left(1 - \exp\left[-\beta_1 \tilde{r} \theta^\epsilon\right]\right)
                   \left(1 - \exp\left[\beta_2 \tilde{r} (\theta-\pi/2)\right]\right)
\end{equation}
Here $\tilde{r} = r/L$ is a nondimensional radius based on a length scale $L = 1.09 \times 10^{10}$ cm and $\tilde{R} = R/L$.  The parameters we use here are as follows: 
$\psi_0 = 4.32 \times 10^{13}$ cm$^2$ s$^{-1}$, $\theta_0 = 0$, $r_b = 0.69$, $r_0 = (\tilde{R}-r_b)/5$, $\Gamma = 3$, $\beta_1 = 0.1$, $\beta_2 = 0.3$, and $\epsilon = 2 + 10^{-8}$.  The resulting profile is illustrated in Fig.\ \ref{fig:ingredients}$b$.
The nondimensional density stratification is given by
\begin{equation}\label{rho}
\tilde{\rho} = \left(\frac{R}{r} - 0.97\right)^n
\end{equation}
with $n = 1.5$.

We use a two-step diffusivity profile, after \cite{dikpa07}
\begin{equation}\label{etadg}
\eta = \eta_c 
+ \frac{\eta_{mid}}{2}
\left[1 + \mbox{erf}\left(2 ~ \frac{r - r_{da}}{d_a}\right)\right]
+ \frac{\eta_{top}}{2}
\left[1 + \mbox{erf}\left(2 ~ \frac{r - r_{db}}{d_b}\right)\right]  ~~~,
\end{equation}
where $\eta_c = 10^9$ cm$^2$ s$^{-1}$, $\eta_{mid} = 5 \times 10^{10}$ cm$^2$ s$^{-1}$, 
$r_{da} = 0.725 R$, $d_{a} = 0.0125 R$, $r_{db} = 0.956 R$, and $d_{b} = 0.05 R$.
We consider two values of the turbulent diffusivity in the surface layers:
$\eta_{top} = 3 \times 10^{12}$ cm$^2$ s$^{-1}$ as in \cite{dikpa07}, as well
as a lower value of $\eta_{top} = 10^{12}$ cm$^2$ s$^{-1}$.  The resulting
profile is illustrated in Fig.\ \ref{fig:ingredients}$c$.

Note that the relatively low value of $\eta$ in the mid CZ, 
$5 \times 10^{10}$ cm$^2$ s$^{-1}$ places our simulations in the so-called
advection-dominated regime in which the meridional flow dominates over
turbulent diffusion for transporting poloidal magnetic flux across the 
CZ \citep[e.g.][]{yeate08,dikpa09,charb10}.  
Though this serves as a good test case, several recent studies have 
suggested that the diffusion-dominated regime may be more realistic,
based on the correlation of the polar field at solar minimum with the 
strength of the next cycle \citep{jiang07}, the Waldmeier effect \citep{karak11}, 
and the efficiency of turbulent transport inferred from helioseismic 
measurements \citep{miesc12b}.  We will consider the diffusion-dominated 
regime in future work.

When solving eq.\ (\ref{eq:indy}), we express the magnetic field 
in terms of poloidal and toroidal components
$\BB = \curl \left(A \uvr\right) + \curl \curl \left(C \uvr\right)$
and we impose radial-field boundary conditions at the top
($A = \pd C/\pd r = 0$ at $r=R$) and perfectly conducting
boundary conditions at the bottom ($\pd A/\pd r = C = 0$ at
$r = r_b$).  All simulations are initiated at $t=0$ with a dipolar
seed field that grows and saturates as described in 
sections \ref{sec:spotmaker}.

\subsection{SpotMaker}
\label{sec:spotmaker}

The spot deposition algorithm is described in MD14 and we refer the reader to that paper for further details.  We call it SpotMaker and it's purpose is to place bipolar magnetic regions (BMRs) on the solar surface in response to the dynamo-generated magnetic field.  The subsequent evolution of these BMRs due to differential rotation, meridional circulation, and turbulent diffusion naturally generates a mean poloidal field as originally described by \cite{babco61} and \cite{leigh64}.  SpotMaker can be regarded as a 3D generalization of the double-ring algorithm developed by \cite{durne97}, \cite{nandy01} and \cite{munoz10}.  A similar axisymmetric BMR formulation was also used by \cite{jiang13} when assimilating sunspot data into a 2D FTD model through the subsurface extrapolation of surface fields.

In SpotMaker, BMRs are placed on the surface based on a spot-producing toroidal field $B^*(\theta,\phi,t)$, which is obtained from $B_\phi(r,\theta,\phi,t)$ by first averaging over radius in the lower CZ (0.70-0.71 $R$) and then applying a mask that excludes latitudes above 40$^\circ$; see eqs.\ (2) and (3) in MD14.  We refer to the maximum value of $B^*(\theta,\phi,t)$ in the northern and southern hemispheres respectively as $B^*_n(t)$ and $B^*_s(t)$.  In order for a BMR to be produced in the northern hemisphere (NH), $B^*_n(t)$ must exceed a threshold value, here taken to be 1 kG.  Similarly for the southern hemisphere (SH) and $B^*_s(t)$.  The latitude and longitude of each BMR is chosen randomly from all horizontal grid points where $B^*(\theta,\phi,t)$ exceeds $B^*_n(t)/2$ or $B^*_s(t)/2$, depending on the hemisphere.

\begin{figure}
\begin{center}
\includegraphics*[width=\linewidth]{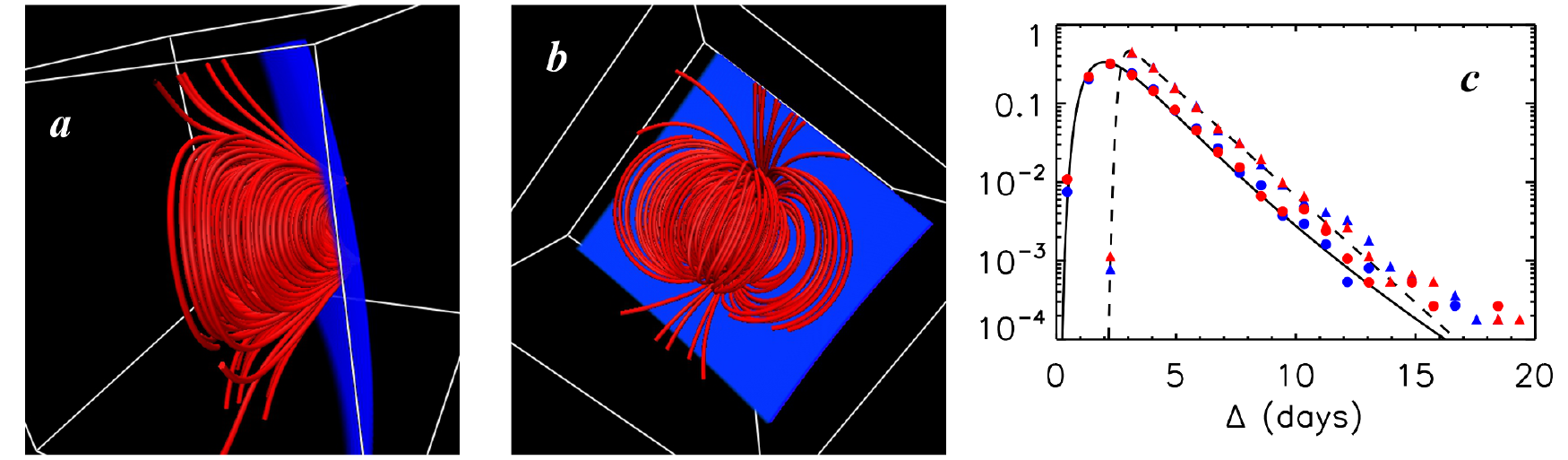}
\end{center}
\caption{($a$,$b$) Subsurface structure of a BMR produced by SpotMaker and ($c$) time lag pdf of emergence events.  The volume rendering in frames ($a$) and ($b$) shows magnetic field lines below the solar surface (red) from two different vantage points, ($a$) east of the BMR looking west and ($b$) underneath the BMR, looking up.  The blue surface represents the surface of the Sun ($r=R$).  The curves in frame ($c$) represent the log-normal pdf (solid line, $\tau_p = 2$ days, $\tau_s = 3$ days) and the sawtooth pdf (dashed line) as given by eqs.\ (\ref{eq:lognormal}) and (\ref{eq:sawtooth}).  Plot symbols represent normalized histograms of the actual BMR lag times in Cases L1 (circles) and S1 (triangles), for the northern (blue) and southern (red) hemispheres.\label{fig:spotmaker}}
\end{figure}

In addition to the threshold field strength, the timing of BMR creation is governed by a time delay probability density function (pdf) $P(\Delta)$, where $\Delta$ is the time that has elapsed since the last BMR creation in each hemisphere.  For example, suppose that a BMR appeared in the NH at time $t_0$.  The timing of the next emergence event (BMR creation) in the NH is then given by $t_1 = t_0 + \Delta_n$, where $\Delta_n$ is chosen randomly based on the time delay pdf $P(\Delta)$.  Similar records are kept independently for the SH, so the emergence events in each hemisphere are asynchronous.

We consider two forms for $P(\Delta)$, illustrated in Fig.\ \ref{fig:spotmaker}$c$.  The first is a lognormal pdf given by
\begin{equation}\label{eq:lognormal}
P_{ln}(\Delta) = \frac{1}{\Delta \sigma \sqrt{2\pi}} ~ 
\exp \left[- \frac{(\ln \Delta - \mu)^2}{2 \sigma^2} \right]  ~~~.
\end{equation}
In practice we specify the mean and mode of the distribution, $\tau_s$ and $\tau_p$ and compute $\sigma^2 = (2/3) \left[\ln(\tau_s) - \ln(\tau_p)\right]$ and $\mu = \ln \tau_p + \sigma^2$.  This is similar to the time delay pdf used by MD14 but there it was implemented somewhat differently, based on the cumulative pdf.
We also consider a sawtooth pdf that can be approximated as an asymmetric stretched exponential as follows:
\begin{equation}\label{eq:sawtooth}
P_s(\Delta) = P_0 \exp\left[ - \frac{\vert \Delta - \Delta_0\vert^{n_\pm}}{\sigma_\pm} \right]  ~~~~,
\end{equation}
where $n_\pm$ and $\sigma_\pm$ have different values depending on the sign of $\Delta - \Delta_0$.  Here we use $\Delta_0 = $3.1 days, $n_{+} = 1$, $\sigma_{+} = 1.6$ days, $n_{-} = 4$, $\sigma_{-} = 0.6$ days.  $P_0$ is a normalization factor ensuring that $\int_0^\infty P_s(\Delta) d\Delta = 1$.   

Once the timing and location of a BMR is determined, the next step is to specify its spatial structure.  This is done by defining a pair of spots on the surface, each specified by a radial magnetic field with circular cross section and a polynomial profile;  see eq.\ (5) in MD14.  Distances on the solar surface are computed using the haversine formula.  The distance between the two spots of a BMR is given by $s_a r_s$, where $r_s$ is the radius of each spot (see below), and the trailing spot (in the sense of rotational motion) is displaced poleward relative to the leading spot at an angle that is given by Joy's Law; $\delta = 32^\circ.1 \cos\theta$ \citep{stenf12}. 

The subsurface structure of each spot pair is determined by means of a potential field extrapolation below the surface; see eq.\ (7) in MD14 and Fig.\ \ref{fig:spotmaker}$a$,$b$.  We realize that this is a gross approximation to the true subsurface structure of active regions but it serves to localize the BL poloidal field generation to the upper CZ, as in previous 2D FTD models
\citep{dikpa09,charb10,munoz10,karak14}.  The boundary conditions for this subsurface extrapolation ensure that the radial field is equal to the imposed BMR field at the surface ($r=R$) and vanishes below a specified penetration radius, $r = r_p$.  The order of the Laplacian precludes further boundary conditions so the horizontal field is not necessarily zero at the surface.  This nominally violates the upper boundary condition but this violation is quickly corrected within a few time steps with the help of explicit and implicit (numerical) diffusion which quickly make the BMR field radial at the surface.

\begin{figure}
\begin{center}
\includegraphics*[width=\linewidth]{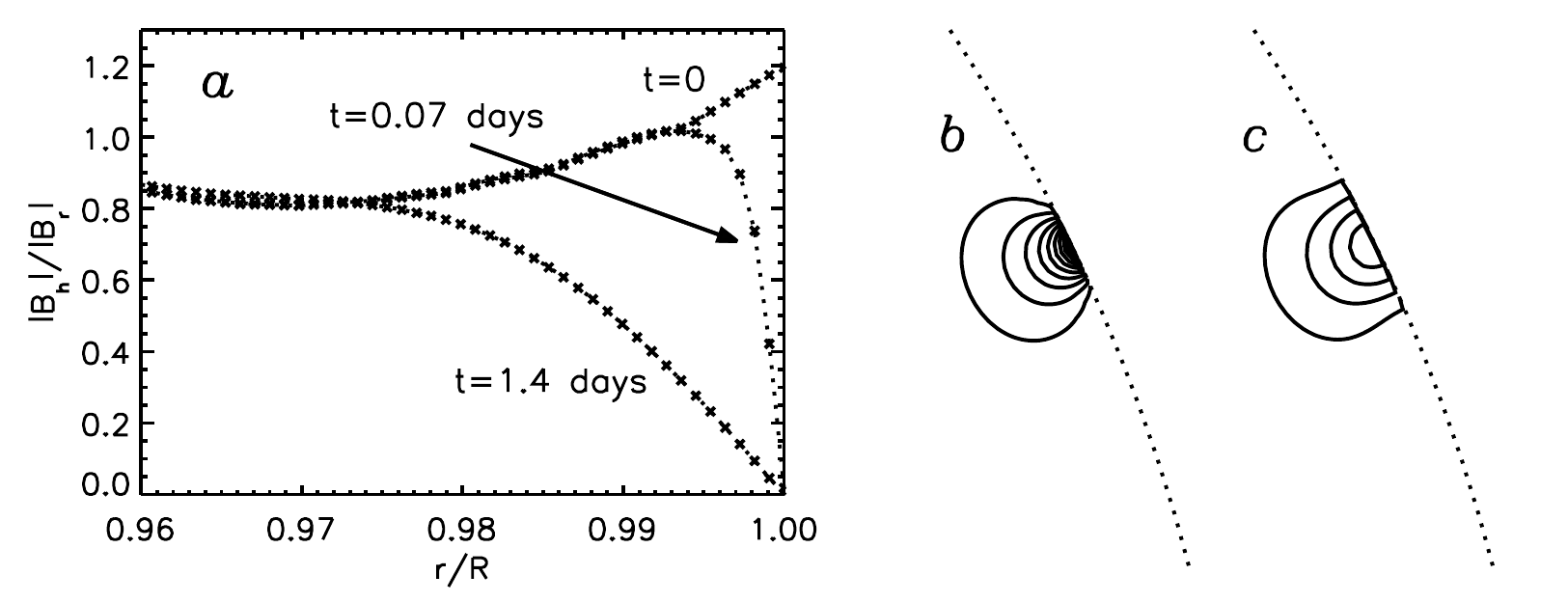}
\end{center}
\caption{Adjustment of the mean ($m=0$) field in a simulation that was initiated with a single BMR.   ($a$) Ratio of horizontal to vertical field strength $\vert B_h \vert / \vert B_r\vert$ as a function of radius near the upper boundary.  $\vert B_h \vert$ and $\vert B_r\vert$ are each averaged over the northern hemisphere before computing this ratio.  The three curves correspond to the initial field ($t=0$), the field after one time step ($t = 0.07$ days) and the field after 20 time steps ($t = 1.4$ days), as indicated.  Crosses represent the radial grid points.  Frames ($b$) and ($c$) show the structure of the mean poloidal field at $t=0$ and after 20 time steps ($t=1.4$ days) respectively.\label{fig:Badjust}}
\end{figure}

This initial adjustment is illustrated in Fig.\ \ref{fig:Badjust} for the mean ($m=0$) component of $\BB$ because this is easiest to visualize.  This is from a simulation that was initiated with a single BMR at a latitude of 25$^\circ$ and a penetration depth of $r_p = 0.90 R$.  The initial magnetic field is identical to that shown in Fig.\ \ref{fig:spotmaker}$a$,$b$ but here the simulation is stepped forward in time, following the evolution of the BMR as it is subject to differential rotation, meridional circulation, diffusion, and the boundary conditions.  This latitude gives it a tilt angle of 13.6$^\circ$ according to the Joy's law expression given above, so it starts out with a nonzero mean poloidal field component (Fig.\ \ref{fig:Badjust}$b$). As noted above, the initial field does not satisfy the radial field upper boundary condition (Fig.\ \ref{fig:Badjust}$a$).  However, after one time step ($t = 0.07$ days), the boundary condition is applied and the horizontal field goes to zero at the surface (Fig.\ \ref{fig:Badjust}$a$).  The turbulent diffusion and the semi-implicit timestepping ensures a smooth transition to the non-zero horizontal field in the interior.  By 20 time steps ($t=1.4$ days), the adjustment is complete, transitioning to a radial field for $r > 0.97 R$ (spanning $\sim$ 30 grid points; Fig.\ \ref{fig:Badjust}$a$,$c$).  Note that the time step used for this simulation (0.07 days) is the same as that used for all the simulations reported in this paper.  The adjustment is similar for the other field components ($m > 0$).

The magnetic flux in each BMR is given by
\begin{equation}\label{eq:flux}
\Phi_s = 2 \Phi_0 ~ \frac{\vert \hat{B}(\theta_s,\phi_s,t_s) \vert}{B_q}
~ \frac{10^{23}}{1 + \left(\hat{B}(\theta_s,\phi_s,t_s)/B_q\right)^2}  \approx B_s r_s^2  ~~~.
\end{equation}
Here $\hat{B}(\theta_s,\phi_s,t_s)$ is the same as $B^*(\theta_s,\phi_s,t_s)$ but without the mask  that suppresses high latitudes (see above).  In short, it is the value of $B_\phi(r,\theta,\phi,t)$ taken at the location and time of the BMR ($\theta_s$, $\phi_s$, $t_s$), averaged over a thin radial region near the base of the CZ (0.70-0.71 $R$).  $B_q$ is a quenching field strength that governs the saturation of the dynamo.  Here we use $B_q = 10^5$ G.  The parameter $\Phi_0$ regulates the flux budget of each BMR and can be increased to achieve supercritical solutions (see section \ref{sec:supercritical}).  The normalization in eq.\ (\ref{eq:flux}) is defined such that for $\Phi_0 = 1$, the strongest BMRs ($\hat{B} = B_q$) have a flux of $10^{23}$ Mx, roughly consistent with solar observations.  Currently STABLE does not account for the depletion of toroidal flux in the lower CZ/tachocline when a new BMR is created.  We will add this capability in future versions of the model.

The radius of each spot is determined by its flux content as $r_s = (\Phi_s / B_s)^{1/2}$, where $B_s$ is set to 3 kG.  Note that we have neglected a factor of order unity ($0.3 \pi$) in the effective spot area ($\int B_r dA / B_s$).  Also, we impose a minimum size of $r_s = $16 Mm to ensure that all BMRs are well resolved and a maximum size of $r_s = 41$ Mm for the largest spots.  If the above calculation for $r_s$ falls outside of these bounds, then $r_s$ is set to its maximum or minimum value and $B_s$ is readjusted to give the requisite flux: $B_s = r_s^{-2} \Phi_s$.

\subsection{A Note on Flux Depletion}\label{sec:fluxdep}

We close this section with a note about flux depletion, which is a more sublte issue than it may first appear.  To illustrate the problem, consider the process of flux emergence, beginning with a coherent toroidal field near the base of the CZ.  For simplicity we can assume that this initial toroidal field is an axisymmetric flux tube with a flux equal to $\Phi_s$ but relaxing this assumption does not change the essential arguments.  Now assume that this flux tube rises and emerges, forming a BMR at the surface with an approximately east-west orientation.  Now define the toroidal flux through any meridional plane as
\begin{equation}
\Phi_{cut}(\phi,r_o,t) = \int_0^\pi \int_0^{r_o} B_\phi(r,\theta,\phi,t) ~ r dr d\theta ~~~,
\end{equation}
where $r_o$ is the outer radius of the domain in question.  If we set $r_o=\infty$ and if we neglect magnetic diffusion in the flux emergence process described above, then $\Phi_{cut}(\phi,\infty,t)$ will be independent of $\phi$ and $t$.  This follows from the topological properties of $\BB$.  Now define the emergence time as $t_e$ and consider a longitude $\phi_b$ that bisects the BMR, lying  between the two polarities.  Again, from the topological properties of $\BB$, we can say that the contribution to $\Phi_{cut}(\phi_b,\infty,t > t_e)$ from the emergent field, $r > R$, must be equal to $\Phi_s$.  In other words, the toroidal flux at longitude $\phi_b$ in the solar interior $r < R$ is depleted by an amount $\Phi_s$ due to the emergence; $\Phi_{cut}(\phi_b,R,t > t_e) = \Phi_{cut}(\phi_b,R,t < t_e) - \Phi_s$.  

Now consider the mean toroidal magnetic flux threading through the computational domain of the model
\begin{equation}
\Phi_{mean}(t) = \int_0^{2\pi} \int_0^\pi \int_{r_b}^{R} B_\phi(r,\theta,\phi,t) ~ r dr d\theta d\phi ~~~.
\end{equation}
This will also be depleted by the emergence process.  However, the amount it will be depleted will depend on the amount of the tube that has emerged, such that
\begin{equation}\label{eq:fluxdep}
\Phi_{mean}(t < t_e) - \Phi_s < \Phi_{mean}(t > t_e) < \Phi_{mean}(t < t_e)  ~~~.
\end{equation}
This is the nature of the flux depletion problem.  As stated above, our SpotMaker algorithm can be regarded as a 3D generalization of the axisymmetric double-ring algorithm described by \cite{nandy01} and \cite{munoz10}.  In those and other papers, the authors take into account flux depletion by subtracting the BMR flux $\Phi_s$ from the mean toroidal flux near the base of the CZ.  However, it should be noted that this corresponds to the maximum depletion limit given by eq.\ (\ref{eq:fluxdep}), $\Phi_{mean}(t > t_e) = \Phi_{mean}(t < t_e) - \Phi_s$.  This limit would only strictly apply if the entire toroidal flux tube were to pass through the solar surface.  Since the presence of a BMR requires some portion of the tube to remain below the solar surface, this limit over-estimates the mean flux loss from an emergence event.  So, it should be regarded as a conservative upper limit of flux depletion.

To highlight this point further, we go back to the orginal configuration of an axisymmetric toroidal flux tube with flux $\Phi_s$.  Now assume that during flux emergence, only a small segment of this tube, spanning a longitudinal range $\Delta \phi$, rises and exits the CZ, leaving the rest of the tube below the surface.  Then the mean toroidal flux $\Phi_{mean}$ will be depleted by an amount that is approximately equal to $\Phi_s \Delta \phi / 2\pi$.  Observations indicate that most BMRs have a longitudinal extent of less than 10$^\circ$ \citep{upton14a}, implying a mean flux depletion of less than 3\% of $\Phi_s$.  Though the depletion of mean flux from the tachocline is likely more than this, much of this mean flux would be redistributed throughout the CZ and, given our incomplete understanding of flux emergence, we currently have no reliable prescription for how best to redistribute this flux.

SpotMaker effectively selects the lower limit of mean flux depletion, $\Phi_{mean}(t > t_e) = \Phi_{mean}(t < t_e)$, which is valid if only a small segment of the tube emerges ($\Delta \phi << 2\pi$).  Though our imposed BMR field clearly has a subsurface east-west component ($B_\phi \neq 0$; see Fig.\ \ref{fig:spotmaker}), the topological nature of the field is poloidal, so it does not change the mean toroidal field $\left<B_\phi\right>$.  However, at the longitude of emergence, the mean flux is indeed depleted in the sense that $\Phi_{cut}(\phi_b,R,t > t_e) \approx \Phi_{cut}(\phi_b,R,t < t_e) - \Phi_s$, as described above.  Unlike previous 2D FTD models, this flux depletion occurs in the upper CZ rather than the lower CZ/tachocline.

A related question is whether or not the flux emergence process conserves magnetic energy.  The current SpotMaker algorithm as laid out here does not; the placement of a BMR increases the magnetic energy.  If the progenitor toroidal field is re-established quickly by differential rotation, this approximation may be justified.  The lifting and twisting of the field during flux emergence can also increase magnetic energy, as in alternative formulations of the Babcock-Leighton mechanism that are based on a local or non-local $\alpha$-effect (the kinematic $\alpha$-effect also does not conserve energy).  Still, a more careful treatment of the magnetic flux budget and energetics is certainly warranted and will be pursued in the future.  The idealized algorithm presented here should be considered as only a starting point, providing a baseline for comparison as we and others develop more sophisticated flux emergence models.


\section{Code Verification}
\label{sec:verification}

The ASH code has already been verified by comparing it to three other independent codes on carefully selected hydrodynamic and MHD simulations of global convection \citep{jones11}.  Here we wish to verify the kinematic version of ASH that has been used as a dynamical core for STABLE.  As mentioned in \S\ref{sec:intro} and \S\ref{sec:STABLE}, in the special case of kinematic, axisymmetric flow fields, the axisymmetric ($m = 0$) component of the induction equation (\ref{eq:indy}) decouples from the non-axisymmetric components so it behaves like a 2D FTD model.  Thus, though STABLE simulations are explicitly 3D, we can legitimately verify the mean ($m=0$) field components by comparing them with an equivalent 2D model.  

\begin{figure}
\begin{center}
\includegraphics*[width=\linewidth]{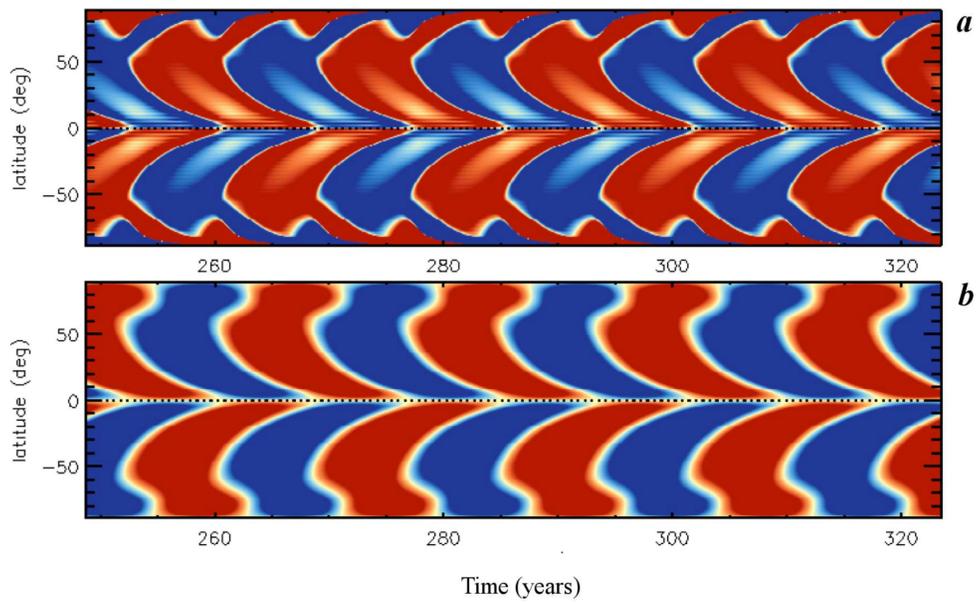}
\end{center}
\caption{Butterfly diagram for the benchmark case SC$^\prime$ of \cite{jouve08}.
($a$) Mean radial field $\left<B_r\right>$ at the surface ($r = R$) as a function of latitude and time. Blue and red denote inward and outward polarity respectively.
($b$) Mean toroidal field $\left<B_\phi\right>$ near the base of the convection zone ($r = 0.70 R$; blue westward, red eastward). Compare with Fig.\ 14 in \cite{jouve08}.\label{fig:bmbfly}}
\end{figure}

To perform this verification, we consider the 2D FTD benchmark simulations defined by \cite{jouve08}.  Specifically, we seek to reproduce their case SC$^\prime$, which is an FTD model in which the source term for the mean poloidal field is a non-local $\alpha$-effect intended to mimic the BL mechanism.  Thus, for the purpose of verification, we replace the BMR deposition algorithm described in \ref{sec:spotmaker} with an explicit poloidal source term as defined in eq.\ (18) of \cite{jouve08}:
\begin{eqnarray}\label{eq:source}
&&\nonumber S(r,\theta,t,B_\phi) = \frac{1}{2}\left[1 + {\rm erf}(\frac{r-r_{bm}}{d_{bm}})\right]\left[1 - {\rm erf} (\frac{r-R}{d_{bm}})\right]\\&&
\left[1 + \left(\frac{B_\phi(r_c,\theta,t)}{B_0}\right)^2\right]^{-1}\cos\theta\sin\theta ~ B_\phi(r_c,\theta,t) ~~~.
\end{eqnarray}
where $r_{bm}= 0.95 R$, $d_{bm} = 0.01 R$ and $r_c = 0.7 R$, as above.  This is implemented by adding a term to the right-hand-side of eq.\ (\ref{eq:indy}) of the form $\curl \left(S \uvp\right)$.  Note that the presence of the quenching term involving $B_0$ provides a saturation mechanism for the dynamo, preventing the magnetic energy from growing exponentially without bound.  Here we use $B_0 = 2 \times 10^5$ G.  We also replace the velocity field and turbulent diffusion in eqs.\ (\ref{omega1}), (\ref{psim}), and (\ref{etadg}) with the corresponding expressions in eqs.\ (13), (19), and (14) of \cite{jouve08}.  This yields a single meridional circulation cell per hemisphere, qualitatively similar to Fig.\ \ref{fig:ingredients}$b$, directed poleward at the surface, equatorward near the base of the CZ, and vanishing at the bottom boundary $r_b$ = 0.65.  Thus, it extends a little below the tachocline, which is centered at $r_c = 0.7 R$.  The boundary conditions are as described in \S\ref{sec:kin}; perfectly conducting at the bottom of the shell and radial field at the top.  For this benchmark, we use a resolution of $N_\theta, N_\phi, N_r$ = 128, 256, 100.

\begin{figure}
\begin{center}
\includegraphics*[width=\linewidth]{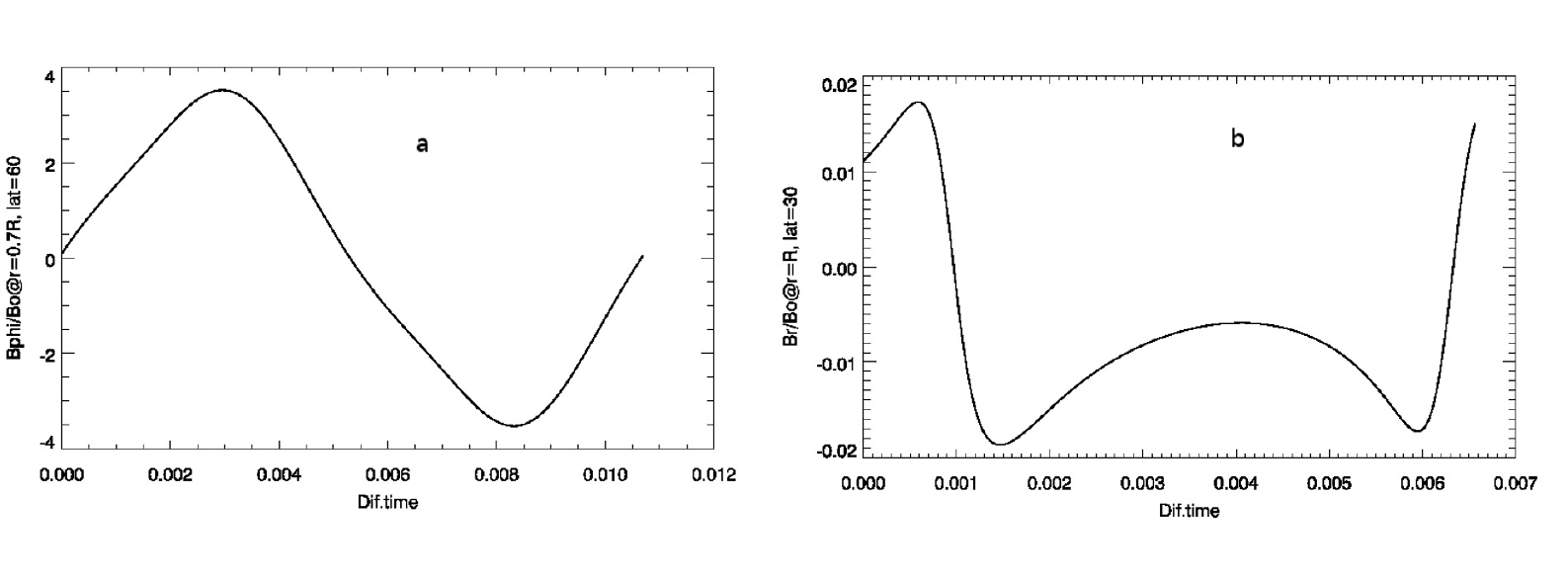}
\end{center}
\caption{Quantitative results for benchmark case SC$^\prime$ of \cite{jouve08}.
($a$) Mean toroidal field $\left<B_\phi\right>$, normalized by $B_0$, at $r = 0.7 R$ 
and latitude = $60^\circ$.  The abscissa is the nondimensional diffusion time 
$\tau_d = (t-t_0) \eta_0 R^{-2}$, where $\eta_0 = 10^{11}$ cm$^2$ s$^{-1}$ and 
$t_0$ is chosen such that the phase of the cycle corresponds with Fig.\ 15 
in \cite{jouve08}.  ($b$) Normalized mean radial field $\left<B_r\right>/B_0$
as a function of $\tau_d$ at $r = R$ and latitude = $30^\circ$.  
Compare with Fig.\ 15 in \cite{jouve08}.\label{fig:compare2}}
\end{figure}

Results are shown in Figs.\ \ref{fig:bmbfly} and \ref{fig:compare2}.  These agree well both qualitatively and quantitatively with the results presented in \cite{jouve08} for eight independent codes.  Note that this simulation is performed in 3D but the fields remain axisymmetric due to the nature of the poloidal source; ME$_{nax}$ as defined in \S\ref{sec:overview} is zero.  We consider this a successful verification of the kinematic STABLE model.

\section{A Representative Dynamo Simulation}
\label{sec:flagship}

Illustrative results from a typical STABLE dynamo simulation are shown in Figures \ref{fig:bfly}--\ref{fig:scalar}.   These are all from Case S1, with parameters summarized in Table 1.  These and other parameters are defined in \S\ref{sec:STABLE} and will be discussed further in \S\ref{sec:supercritical}.  Here we give a general overview of the self-sustained dynamo solutions achieved with STABLE.  Like the other simulations described in \S\ref{sec:supercritical}, this case was done with a resolution of $N_\theta, N_\phi, N_r$ = 512, 1024, 340.

\subsection{Overview of Cycle Characteristics}
\label{sec:overview}

\begin{figure}
\begin{center}
\includegraphics*[width=\linewidth]{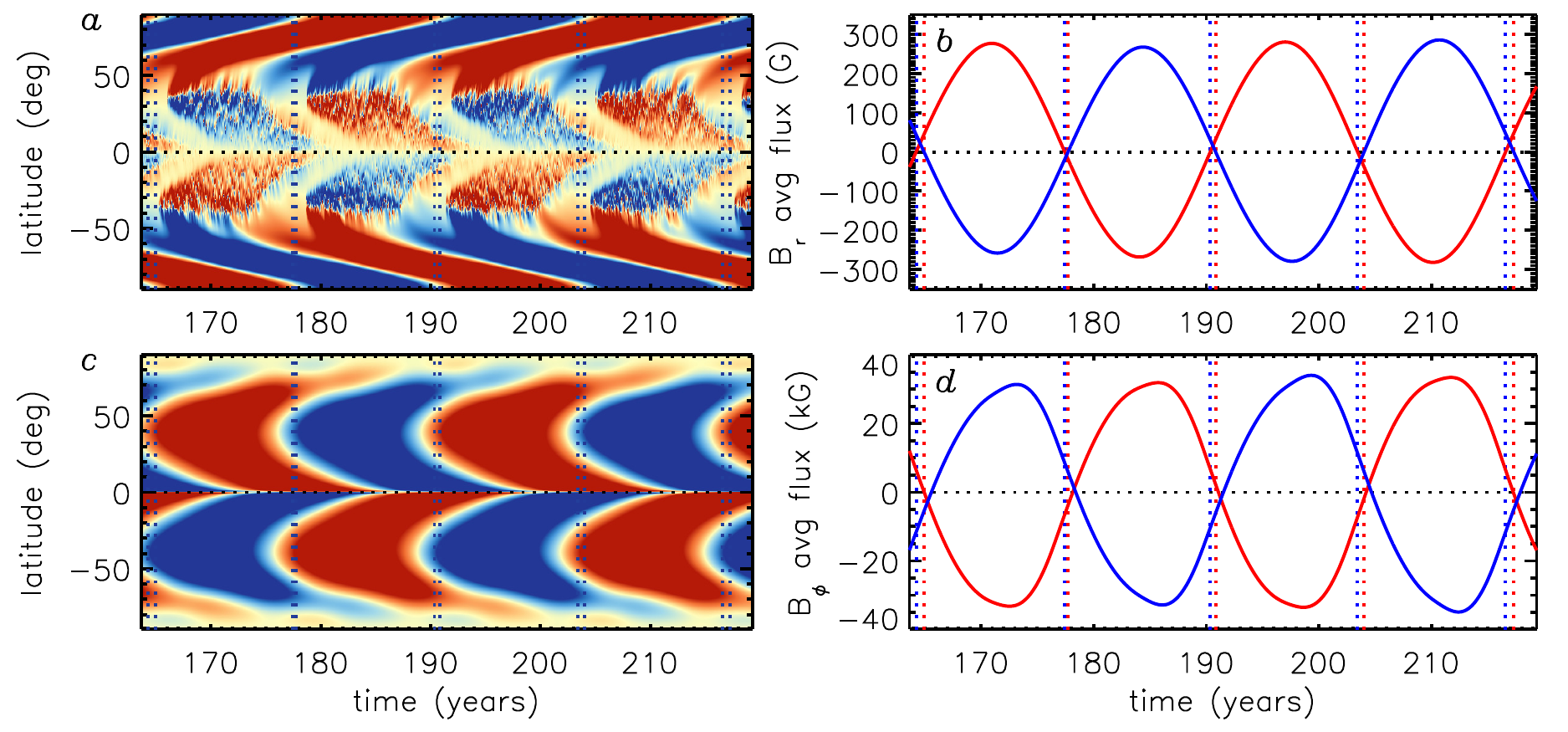}
\end{center}
\caption{Magnetic cycles in Case S1. ($a$) $\left<B_r\right>$ at the surface ($r=R$) as a function of latitude and time, highlighting four magnetic cycles.  Red and blue denote outward and inward field respectively.  Peak amplitudes can exceed 300 G but the color table saturates at $\pm$ 100G. ($b$) $\left<B_r\right>$ averaged over the north (blue) and south (red) polar regions, above a latitude of $\pm$ 85$^\circ$.  Vertical dotted lines in this and all other frames mark polar field reversals in the NH (blue) and SH (red). Frames ($c$) and ($d$) are similar to frames ($a$) and ($b$) but for $\left<B_\phi\right>$ in the lower CZ ($r = 0.71 R$).  However, the averages in ($d$) are over the entire NH (blue) and SH (red), as opposed to just the polar regions.  Red and blue in ($c$) denote eastward and westward field respectively, with a saturation level for the color table of 50 kG.\label{fig:bfly}}
\end{figure}

The magnetic cycles are perhaps best demonstrated by the butterfly diagrams in Figs.\ \ref{fig:bfly}$a$ and $b$.  These show the mean (longitudinally-averaged) radial and toroidal field $\left<B_r\right>$ and $\left<B_\phi\right>$ at $r=R$ and $r=0.71 R$ respectively for the northern hemisphere (NH) and southern hemisphere (SH).  Throughout this paper, angular brackets denote averages over longitude.  Also shown in Fig.\ \ref{fig:bfly} are the time evolution of the ($b$) polar flux and ($d$) mean toroidal magnetic flux in each hemisphere, expressed here as an average field strength.  Polar field reversals are marked with vertical dotted lines.  

\begin{figure}
\begin{center}
\includegraphics*[width=\linewidth]{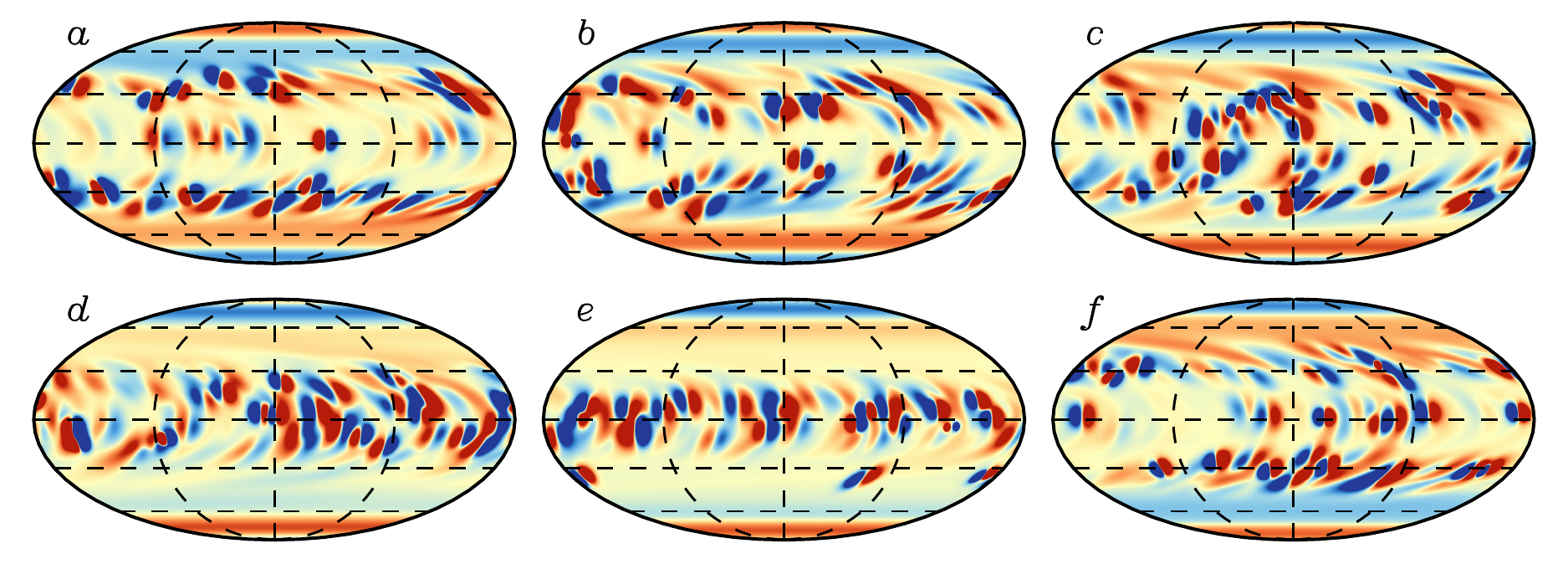}
\end{center}
\caption{Radial magnetic field $B_r$ at the solar surface ($r=R$) in case S1, plotted in Molleweide projection.  Dashed lines denote latitudes of $0^\circ$, $\pm 30^\circ$, and $\pm 60^\circ$.  Six snapshots are shown, spanning one magnetic cycle: $t = $ ($a$) 181.0 ($b$) 183.6, ($c$) 186.2, ($d$) 188.9, ($e$) 191.5, and ($f$) 194.1 years.  Red and blue denote radially outward and inward field respectively, with a saturation level on the color table of $\pm $ 500 G.
\label{fig:ss}}
\end{figure}

As in all advection-dominated FTD models, the equatorward migration of toroidal field at low latitudes in Fig.\ \ref{fig:bfly}$c$ can be attributed to the equatorward meridional circulation near the base of the CZ.  This deep-seated toroidal field is often used as a proxy for sunspots but there is no need for such a proxy with STABLE; BMRs appear at the surface explicitly, and migrate equatorward over the course of the cycle (Fig.\ \ref{fig:bfly}$a$) along with the subsurface $\left<B_\phi\right>$.  The distortion and dispersal of these tilted (Joy's law) BMRs by differential rotation, meridional circulation, and turbulent diffusion gives rise to a poleward migration of trailing flux that reverses the polar fields (Fig.\ \ref{fig:bfly}$a$), as described by the BL mechanism (\S\ref{sec:intro}).  

\begin{figure}
\begin{center}
\includegraphics*[width=\linewidth]{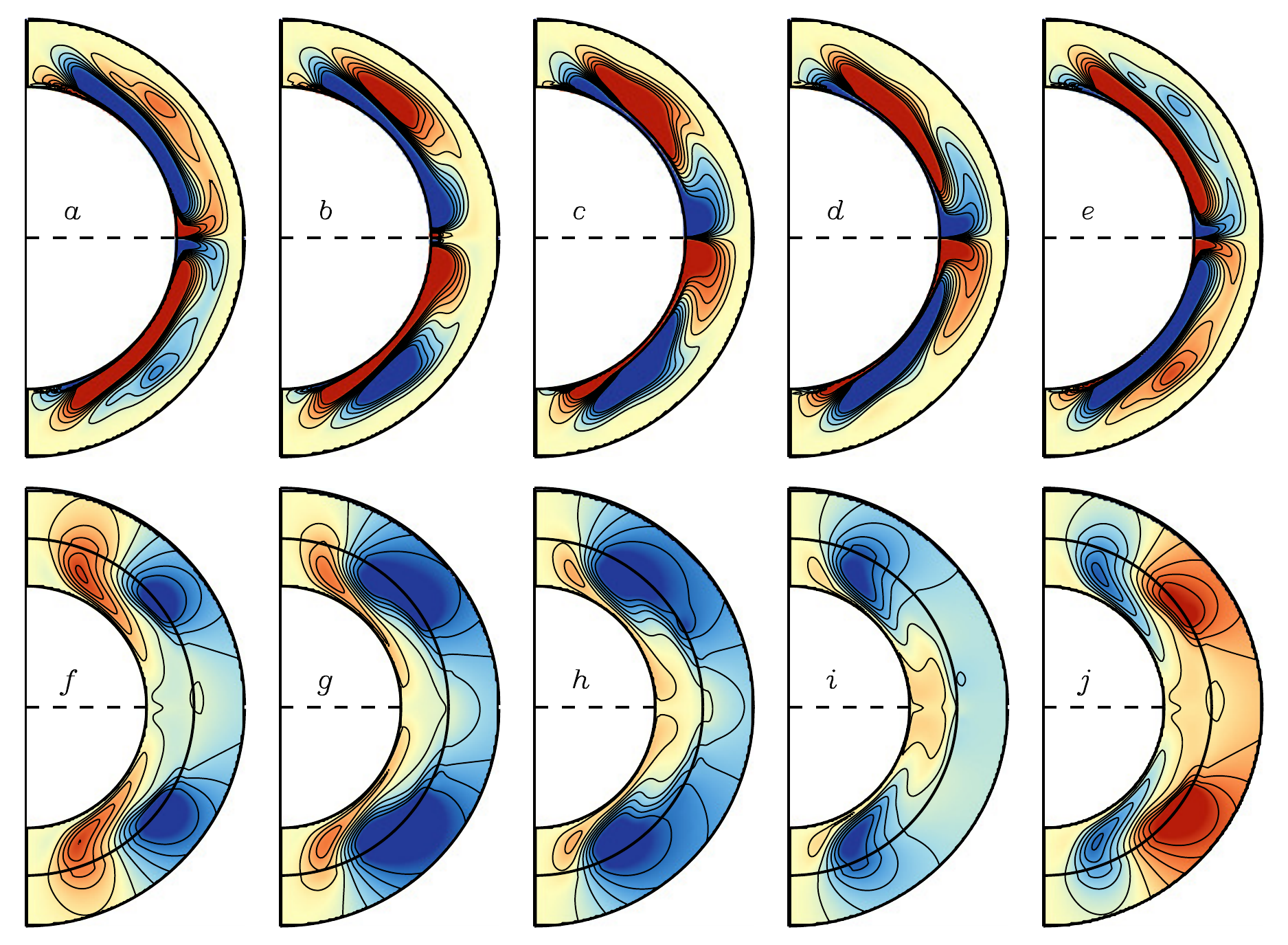}
\end{center}
\caption{Mean ($a$-$e$) toroidal and ($f$-$j$) poloidal magnetic fields in case S1.  Five snapshots are shown, spanning the same magnetic cycle as in Fig.\ \ref{fig:ss}: $t = $ ($a$,$f$) 181.0, ($b$,$g$) 184.3, ($c$,$h$) 187.5, ($d$,$i$) 190.8, ($e$,$j$) 194.1 years. Frames ($a$-$e$) show $\left<B_\phi\right>$, with red and blue indicating eastward and westward field respectively.  Peak field strengths can exceed 100 kG but the color table is clipped at $\pm$ 20 kG.  Frames ($f$-$g$) show the poloidal magnetic potential with a potential-field extrapolation above $r=R$ (to $r=1.25 R$).  Red and blue denote clockwise and counter-clockwise field respectively, with peak values of $\left<B_r\right>$ on the order of 800 G.\label{fig:bmean}}
\end{figure}

A conspicuous shortcoming of this model (to trained eyes) is the time it takes for mid-latitude flux to migrate poleward and reverse the polar fields, which we'll refer to as $\tau_m$.  In this model, $\tau_m$ spans over a decade whereas in the Sun it takes only a few years \citep[e.g.][]{hatha10}.  This can be attributed to the imposed meridional flow, which was originally devised to investigate the impact of high-latitude counter-cells, with diverging flows near the poles \cite{dikpa10b}.  Thus, the speed of the poleward MC at high latitudes is slower than in some other FTD and SFT models \citep{dikpa09,charb10,karak14,upton14a}.  We have confirmed that the use of different MC profiles can substantially reduce $\tau_m$ and thus eliminate this apparent shortcoming of the model.  Results will be presented in a forthcoming paper.  We have also confirmed that $\tau_m$ is insensitive to the magnitude of the turbulent diffusion at the surface, 
$\eta_{top}$.  Simulations with both higher (not shown here) and lower (see cases L2 and L3 in \S\ref{sec:supercritical}) values of $\eta_{top}$ exhibit similar migration time scales $\tau_m$ (see Fig.\ \ref{fig:bflies}).  Another apparent shortcoming of the model is the relatively strong polar fields (Fig.\ \ref{fig:bfly}$b$).  This can be corrected by reducing the parameter $\Phi_0$; see \S\ref{sec:supercritical}.

\begin{figure}
\begin{center}
\includegraphics*[width=\linewidth]{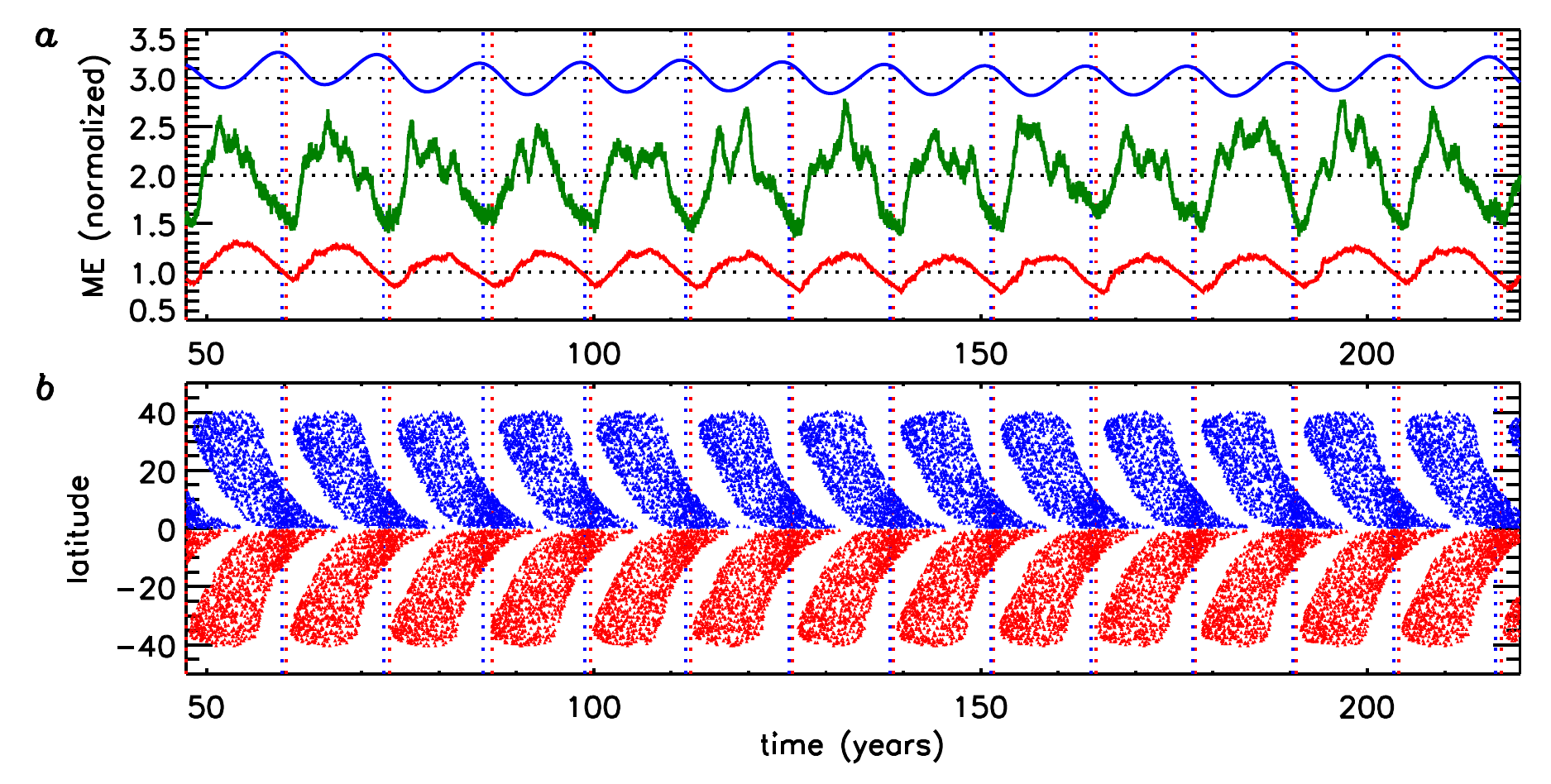}
\end{center}
\caption{($a$) Total magnetic energy in the mean toroidal field, ME$_{tor}$ (blue), the mean poloidal field ME$_{pol}$ (red), and the non-axisymmetric field components ME$_{nax}$ (green) over a time interval spanning 172 years.  All quantities are integrated over the entire computational volume (which spans the entire CZ) and normalized for clarity in plotting.  Normalization factors are $1.15\times 10^{43}$ erg for ME$_{tor}$, $1.24 \times 10^{40}$ erg for ME$_{nax}$, and $1.49 \times 10^{39}$ erg for ME$_{pol}$.  After normalization, we added one to the ME$_{nax}$ curve and two to the ME$_{tor}$ curve in order to plot all three curves with minimal overlap.  Thus, the mean values for the normalized ME$_{tor}$, ME$_{nax}$ and ME$_{pol}$ curves are 3, 2, and 1 respectively, as indicated by the black dotted lines. ($b$) Latitudinal positions of the BMR deposition sites in the NH (blue) and SH (red) over the same time interval.  Over this interval, 14,452 BMRs were introduced in the NH and 14,411 BMRs were introduced in the SH.  Blue and red vertical dotted lines indicated reversal of the polar fields as in Fig.\ \ref{fig:bfly}.
\label{fig:scalar}}
\end{figure}

It is interesting to note that the time evolution of the polar fields (Fig.\ \ref{fig:bfly}$b$) is nearly sinusoidal whereas the toroidal flux is more asymmetric, with a slower rise and faster decay.  Note also the slight phase difference between the northern and southern hemispheres.  Though this often persists for several cycles, the dynamo sporadically re-synchronizes, maintaining a dipolar parity (see Fig.\ \ref{fig:scalar}$a$).  We emphasize again (see \S\ref{sec:STABLE}) that the spot deposition in each hemisphere is asynchronous and that the build-up of the dipole moment is cumulative, with contributions from multiple active regions.  So, the cross-equator cancellation of surface $B_r$ that regulates the polar field strength \citep{camer13} occurs only in an integrated sense, involving residual flux from many BMRs as in the Sun.  This is the origin of the north-south asymmetry.

Figure \ref{fig:ss} shows the evolution of the surface fields over the course of a magnetic cycle.  A close look at each of these snapshots reveals multiple BMRs, in various stages of evolution.  Localized, newly formed BMRs obey Joy's Law (increasing tilt with latitude; see \S\ref{sec:spotmaker}) by construction and Hale's law (oppose orientation in the NH and SH) by virtue of the dipolar nature of the dynamo mode.  Axisymmetric bands of radial field at high latitude arise from the distortion and dispersal of tilted BMRs and they migrate poleward due mainly to the MC.  A brief cycle overlap can be discerned in Fig.\ \ref{fig:ss}$e$, which shows several BMRs with positive (red) leading polarity at a latitude of about -35$^\circ$ coexisting with several other BMRs near the equator with negative (blue) leading polarity.  Note also the asymmetry apparent in this same Figure: spots for the new cycle appear at mid-latitudes in the SH slightly before they appear in the NH (see also Fig.\ \ref{fig:bfly}$a$).  The evolution of the mean (longitudinally-averaged) fields during this same magnetic cycle is shown in Fig.\ \ref{fig:bmean}.

Most of the magnetic energy (over 99\%) is in the mean toroidal field, ME$_{tor}$, which varies by about 15\% over the course of a cycle.  This is demonstrated in Fig.\ \ref{fig:scalar}$a$ (see also Table 1), which shows the evolution of the total magnetic energy in the mean and non-axisymmetric fields.  Interestingly, the minima of the poloidal magnetic energy ME$_{pol}$ do not coincide with the reversals of the polar fields at the surface.  Rather, they occur slightly after.  Meanwhile, the polar field reversals occur near the maxima of ME$_{tor}$, though slightly after, by about 0.5 to 1.5 years.  This is suggestive of solar observations in which the poloidal field reversals occur near sunspot maximum.  However, in the STABLE model, the magnetic energy in BMRs is reflected mainly by the non-axisymmetric field components, ME$_{nax}$, which reaches a (global) minimum as the polar fields at the surface are reversing.   This is also reflected by the butterfly diagram in Fig.\ \ref{fig:bfly}$a$, which suggests that polar fields reverse near a time of minimum sunspot activity.  Thus, the phasing of toroidal and poloidal fields is not in good agreement with solar observations.  However, it is likely that this aspect of the simulations will improve as we implement different MC profiles that more faithfully capture the poleward migration time scale of trailing magnetic flux $\tau_m$.  See the discussion above in connection with Fig.\ \ref{fig:bfly}.

It is also interesting to note that the evolution of ME$_{nax}$ over the course of a cycle is asymmetric, with a fast rise and a slow decline.  This is similar to the observed evolution of the solar sunspot number \citep{hatha10} but it's opposite to the asymmetry noted above with regard to the integrated toroidal flux in Fig.\ \ref{fig:bfly}$d$.  Dynamo models often use the subsurface toroidal flux as a proxy for the sunspot number.  The differences noted here even for an idealized FTD model such as STABLE suggest that this proxy may not be as reliable as is often assumed.  Also, the distribution of the radial field at any instant quickly spreads beyond the emergence sites of BMRs.  This can be seen by comparing the butterfly diagram in Fig.\ \ref{fig:bfly}$a$ with the actual emergence latitudes in Fig.\ \ref{fig:scalar}$b$.  Note that the corresponding longitudinal locations of the emergent BMRs are random.

\subsection{The Role of Surface Fields in Dynamo Operation}
\label{sec:cs}

From the perspective of space weather/space climate forecasting, one of the beneficial aspects of FTD models is the disproportionate role that surface magnetism plays in the operation of the dynamo\footnote{By ``disproportionate'' we do not mean to imply that subsurface fields are not important; they are of course essential to sustain the dynamo.  We merely mean that surface fields appear to play a greater role in the dynamo than might be expected given their contribution to the total magnetic energy in the CZ and tachocline, which is thought to be relatively small.}.  If the main source of poloidal magnetic flux is indeed the BL mechanism, then we can observe this occurring and we can use this information to help forecast future magnetic activity.  More generally, the generation of the toroidal flux in each hemisphere that is responsible for producing BMRs appears to be linked to the shearing and amplification of the observed poloidal flux that passes through the solar surface.  This was demonstrated recently by \cite{camer15}, hereafter CS15.

The analysis performed by CS15 begins by averaging the longitudinal ($\phi$) component of the MHD magnetic induction equation (\ref{eq:indy}) over $\phi$ and integrating it over the NH.  If the velocity field is assumed to be axisymmetric, as it is here, this procedure yields the following expression for the evolution of the mean toroidal flux through the NH
\begin{equation}\label{eq:dphidt}
\frac{d \Phi_{NH}}{dt} = \int_{\cal S} \curl \left[\vv \cross \left<\BB\right> - \eta_t \curl \left<\BB\right>\right] \bdot d\surf 
= \int_{\delta {\cal S}} \left[\vv \cross \left<\BB\right> - \eta_t \curl \left<\BB\right>\right] \bdot d\bell
\end{equation}
where 
\begin{equation}
\Phi_{NH}(t) = \int_0^{\pi/2} \int_{r_1}^{r_2} \left<B_\phi\right> ~ r dr d\theta   
\equiv \int_{\cal S} \left<\BB\right> \bdot d\surf ~~~.
\end{equation}

\begin{figure}
\begin{center}
\includegraphics*[width=\linewidth]{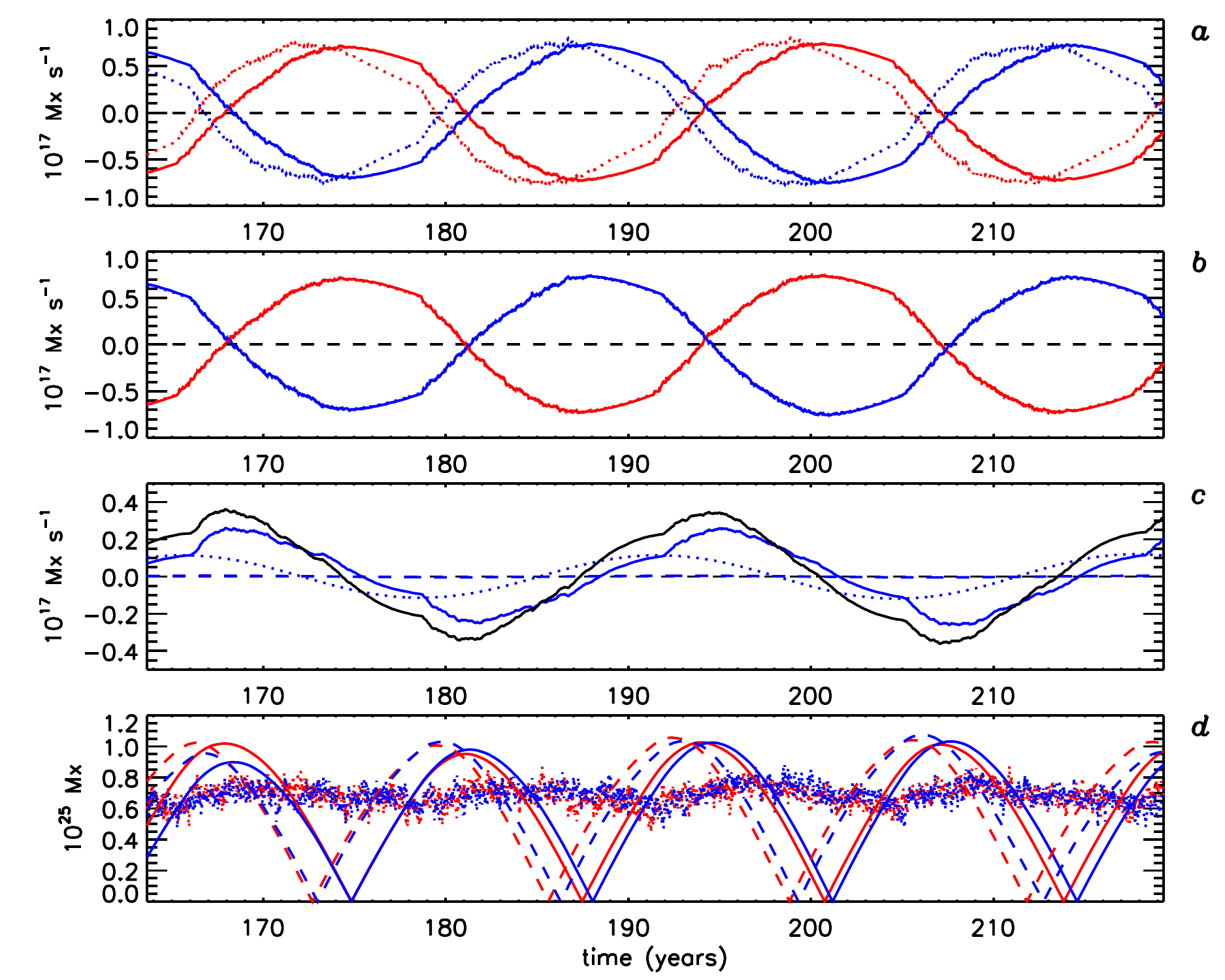}
\end{center}
\caption{($a$) Rate of change of the mean toroidal flux threading the NH $d\Phi_{NH}/dt$ (blue solid line) for a selected time interval in Case S1 spanning four magnetic cycles.  Red solid lines indicate its counterpart in the SH, $d\Phi_{SH}/dt$.  Dotted lines indicate NH (blue) and SH (red) contributions from the surface DR term in eq.\ (\ref{eq:cs}). ($b$) As in ($a$) but including contributions from the diffusive terms in eq.\ (\ref{eq:dphidt}). Now the solid and dotted lines coincide.  ($c$) Breakdown of the diffusive contributions to the line integral in eq.\ (\ref{eq:dphidt}), including contributions from the upper (blue solid line), equatorial (blue dotted line), and polar (blue dashed line) branches of the closed circuit. The sum of these contributions is plotted as a black solid line.  ($d$) Time evolution of the absolute value of the net toroidal flux in each hemisphere, $\vert \Phi_{NH}(t)\vert$ (blue solid line) and $\vert \Phi_{SH}(t)\vert$ (red solid line), together with the predicted evolution obtained by integrating eq.\ (\ref{eq:cs}) over the NH (blue dashed line) and SH (red dashed line).  Dotted lines show the unsigned vertical flux $\vert B_r\vert$ at the surface integrated over the NH (blue) and SH (red).  Compare with Fig.\ 3 in CS15.\label{fig:cs}}
\end{figure}

In eq.\ (\ref{eq:dphidt}), $d\bell$ denotes a differential segment of the closed, clockwise, linear circuit $\delta {\cal S}$ that encircles the NH, proceeding radially outward at the north pole, equatorward at the solar surface, inward at the equator, and poleward just below the base of the CZ.  This astute use of Stokes' theorem links the time evolution of the mean toroidal flux threading through the entire NH to the flows and fields on the boundaries of the CZ.  Furthermore, as CS15 demonstrate, the dominant component of this line integral is the shearing of the radial field by the surface DR, which can be written as follows:
\begin{equation}\label{eq:cs}
S_{NH}^t(t)  = R^2 \int_0^{\pi/2} \left<B_r\right> \left(\Omega(R,\theta,t) - \Omega_{eq}\right) \sin\theta d\theta 
\end{equation}
where $\Omega(r,\theta,t) = \Omega_0 + v_\phi r^{-1} \sin^{-1}\theta$ is the angular velocity.  CS15 chose a reference frame rotating with angular velocity $\Omega_{eq}$, which is the value of $\Omega$ at $r = R$ and $\theta = \pi/2$ (equatorial surface rate).  This choice of reference frame, together with the weak radial dependence of the solar rotation rate at the equator inferred from helioseismology (cf.\ Fig.\ \ref{fig:ingredients}$a$), implies that the contribution from the DR term $(v_\phi \uvp) \cross \left<\BB\right>$ is small for the equatorial portion of the line integral in eq.\ (\ref{eq:dphidt}).  Furthermore, the contribution of this DR term to the inner and polar branches of the line integral vanish if there is no mean flux through those boundaries (as here).  The impenetrable boundary conditions and the symmetry of the MC ($v_\theta = 0$ at the equator) ensure that the MC contributions to the line integral in eq.\ (\ref{eq:dphidt}) also vanish.  Thus, the only terms other than eq.\ (\ref{eq:cs}) that contribute significantly to the line integral in eq.\ (\ref{eq:dphidt}) are the diffusive terms.  Our perfectly-conducting inner BCs preclude diffusion into the deep interior but diffusion along the polar, upper, and equatorial branches of the line integral is in general non-zero.

The evolution of $S_{NH}^t(t)$ and its counterpart, $S_{SH}^t(t)$, are shown in Fig.\ \ref{fig:cs}$a$ (dotted lines).  These are plotted together with the actual time derivative of $\Phi_{NH}(t)$ and $\Phi_{SH}(t)$ (solid lines).  CS15 model the diffusive terms as an effective drag, inducing an exponential decay of $\Phi_{NH,SH}$ in the absence of DR.  This effectively decreases the amplitude of $d \Phi_{NH,SH}/dt$ during the rising phase of a cycle and leads to a negative phase shift such that reversals and extrema occur earlier than they would without the diffusive terms.  However, we find the opposite in our STABLE FTD models.  Namely, the presence of the diffusive terms enhances the amplitude of $d\Phi_{NH,SH}/dt$ during the rising phase, inducing a {\em positive} phase shift such that reversals and extrema occur {\em later} than they would otherwise.  This is demonstrated by Fig.\ \ref{fig:cs}$b$ where the inclusion of the diffusive terms shifts the dotted curve in Fig.\ \ref{fig:cs}$a$ to the right by about 3 years, until it lies on top of the actual $d\Phi_{NH,SH}/dt$. 

To illustrate why this occurs, consider the NH of Fig \ref{fig:bmean}, focusing on the toroidal field evolution in the upper row.  As the flux from the new cycle is building (red), flux from the previous cycle (blue) is pushed equatorward and upward by the MC (Fig.\ \ref{fig:bmean}$b$-$e$).  This causes a decay of the flux from the previous cycle due to diffusion first across the equator and then through the upper boundary (Fig.\ \ref{fig:cs}$c$).  Meanwhile, the diffusion at the poles is negligible.  This diffusive expulsion of the flux from the previous cycle causes the net flux $\Phi_{NH}(t)$ to rise, even after the surface DR stops amplifying the flux through $S_{NH}^t(t)$ (Fig.\ \ref{fig:cs}$a$).  For example, by $t = 194.1$, shown in Fig.\ \ref{fig:bmean}$e$ and $j$, $S_{NH}^t(t)$ has already reversed sign, as the surface DR generates negative toroidal flux from the new, clockwise poloidal field at low latitudes.  Yet, the net toroidal flux is still growing ($d\Phi_{NH}/dt > 0$; see Fig.\ \ref{fig:cs}$a$) due to the selective removal of opposing flux from the previous cycle by diffusion.  This shifts the maximum toward a later time (Fig.\ \ref{fig:cs}$d$).  

Figure \ref{fig:cs}$d$ shows the time evolution of the amplitude of $\Phi_{NH,SH}(t)$ (solid lines) together with the predicted evolution based on integrating equation (\ref{eq:cs}), shown as dashed lines.  The integration constant is chosen such that the zeros of the predicted $\vert\Phi_{NH,SH}(t)\vert$ curves correspond roughly to the derivative extrema in Fig.\ \ref{fig:cs}$a$ (dotted lines).  If we were to include diffusion in the predicted $\vert\Phi_{NH,SH}(t)\vert$ curves, then the result would essentially coincide with the actual $\vert\Phi_{NH,SH}(t)\vert$ curves (solid lines), as in Fig.\ \ref{fig:cs}$b$.

We emphasize again that CS15 have no information on the structure of the field or the role of diffusion below the solar surface.  Instead, they emphasize the importance of the surface DR term, eq.\ (\ref{eq:cs}), which can be computed based on solar observations.  They then model the subsurface diffusion as an exponential decay term that shifts the $\Phi_{NH}(t)$ and $\Phi_{SH}(t)$ curves to the left, toward earlier times.  We confirm the importance of the surface DR term in our idealized FTD model but we find that effect of the subsurface diffusion is instead to shift the $\Phi_{NH}(t)$ and $\Phi_{SH}(t)$ curves to the right, toward later times.  CS15 verify their argument by demonstrating that the predicted peaks of $\vert \Phi_{NH}(t) \vert$ and $\vert \Phi_{SH}(t) \vert$ based on $S_{NH}^t(t)$ and $S_{SH}^t(t)$ correlate with observed maxima in the unsigned surface flux, suggesting that the enhanced surface flux arises from the emergence of greater net subsurface flux.  If the subsurface diffusion were to shift the predicted flux curves to the right instead of to the left as suggested by our model, then this would improve the CS15 correlation, substantiating the CS15 argument (see their Fig.\ 3).  However, that said, in our model, there is little variation of the unsigned surface flux over the course of the cycle and what little variation there is appears to be anti-correlated with the predicted flux (Fig.\ \ref{fig:cs}$d$).  The positive phase shift due to diffusion improves this correlation slightly but there is still about a four-year delay between the maxima in $\vert \Phi_{NH,SH}(t) \vert$ and the peak in the unsigned surface flux.

We close this section by noting that the generation of toroidal flux by the surface DR terms $S_{NH,SH}^t(t)$ generates enough toroidal flux to account for all of the unsigned flux present on the surface.  This is even true for Case S1 (Fig.\ \ref{fig:cs}$d$), in which the flux in BMRs has been artificially enhanced by a factor of five ($\Phi_0 = 5$; see Table 1). CS15 argue that this is also the case for the Sun.

\section{Achieving Self-Sustained Dynamo Action}
\label{sec:supercritical}

A shortcoming of the model presented in \S\ref{sec:flagship} (Case S1) is that the fields are too strong.  For example, the average strength of the polar fields is over 200 G (Fig.\ \ref{fig:bfly}$b$).  The Sun, by comparison, is about 10 G \citep{munoz12}.  This can be attributed in part to the large value of $\Phi_0 = 5$ used to artificially enhance the flux in BMRs as expressed in eq.\ (\ref{eq:flux}).  We used this large value of $\Phi_0$ in order to ensure that the dynamo is supercritical, meaning that sustained dynamo action can occur despite the inhibiting effects of turbulent diffusion. The main focus of this section is to see if we can achieve supercritical solutions with $\Phi_0 = 1$, thus avoiding artificial amplification of the BMR flux.

\begin{figure}
\begin{center}
\includegraphics*[width=\linewidth]{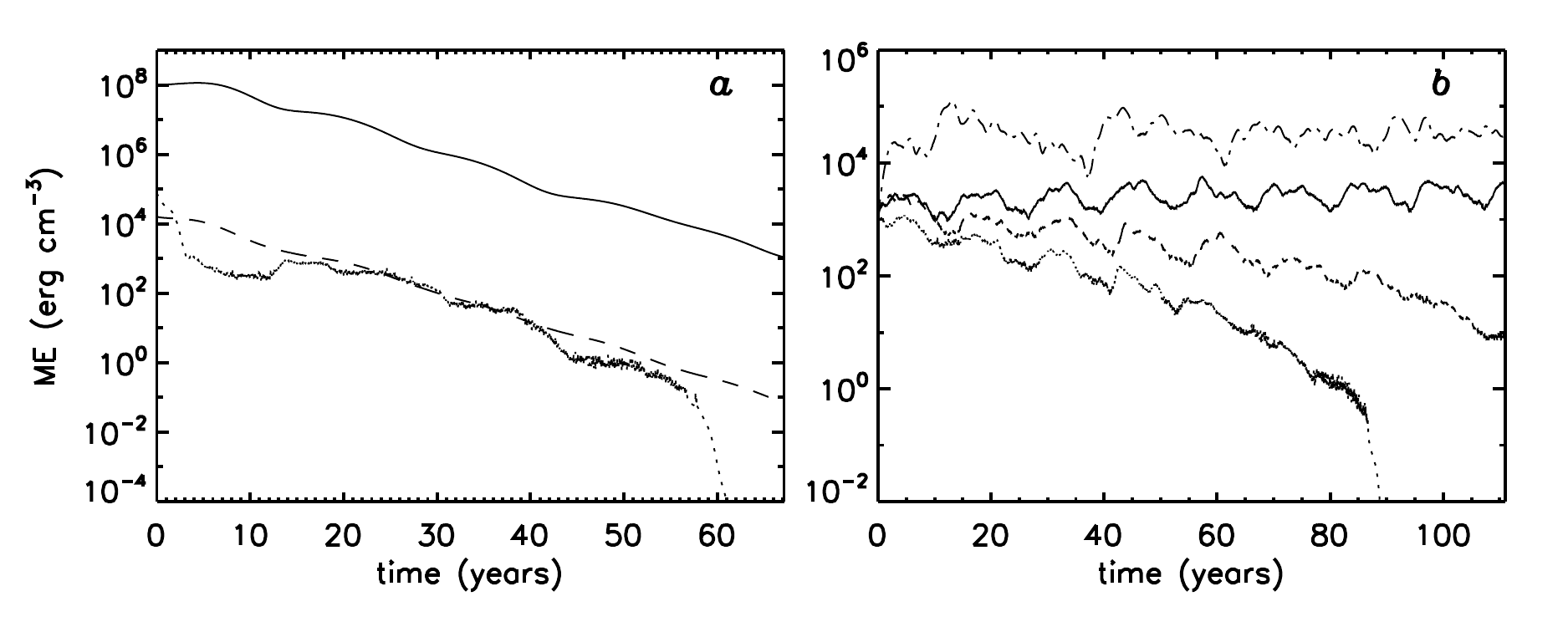}
\end{center}
\caption{($a$) Evolution of the magnetic energy, ME, in Case S6, expressed as an average energy density in erg cm$^{-3}$.  Solid, dashed and dotted lines correspond to the mean toroidal (ME$_{tor}$), mean poloidal (ME$_{pol}$), and non-axisymmetric field components (ME$_{nax}$) respectively.  ($b$) ME$_{nax}$ for cases S8 (dashed line), L1 (dotted line), L2 (solid line), and L3 (dot-dashed line).
\label{fig:metoo}}
\end{figure}

However, before proceeding, we note that the quenching field strength, $B_q$, also contributes to our artificially large polar field values (see eq.\ \ref{eq:flux}).  Here we set $B_q = 10^5$ G and we use the same value for all simulations in order to faciliate a comparison between them.  Since the kinematic induction equation is linear apart from this quenching term, scaling down the value of $B_q$ should also scale down the poloidal, toroidal, and non-axisymmetric field strengths by the same factor.  Thus, it is more meaningful to consider the ratio of toroidal to poloidal field, as argued by \citet{choud03}.  For all of our cases this ratio is similar, at about 85-88 (see Table 1).  Scaling down $B_q$ to achieve a mean polar field strength of about 10G should thus yield a mean toroidal field strength of about 850-880 G.  This is significantly less than the toroidal field strengths inferred from simulations of rising flux tubes, which are in the range 40-100 kG \citep{choud87,calig95,fan09,weber11}.  However, as discussed by \citet{choud03}, the toroidal field could well be highly intermittent, with strong, buoyantly unstable flux tubes embedded in a more diffuse background field.  We also note that a toriodal to poloidal field ratio of under 100 is a common feature of advection-dominated FTD models; see, e.g. \citet{dikpa02}.

If we start with the case described in \S\ref{sec:flagship} (Case S1) and drop the value of $\Phi_0$ from five to unity, the dynamo decays, as shown in Fig.\ \ref{fig:metoo}$a$.  This is Case S6, from a series of simulations summarized in Table 1.  Case S6 was started from the same initial conditions as Case S1 (Fig.\ \ref{fig:scalar}$a$), provided by a progenitor case with higher $\Phi_0$ (not shown).  Though most of the magnetic energy is in the mean toroidal fields, the dynamo cannot operate without the BMRs, which dominate the non-axisymmetric magnetic energy ME$_{nax}$.  As the toroidal field decays due to diffusion, it eventually drops below the threshold field for creating BMRs (see \S\ref{sec:STABLE}).  This causes ME$_{nax}$ to drop rapidly beyond $t \sim $ 57 yrs (Fig.\ \ref{fig:metoo}$a$).  This is the point of no return; once the dynamo ceases to make BMRs, it will continue to decay indefinitely.

The most straightforward way to ensure that the dynamo is supercritical is to artificially boost the flux in BMRs by increasing $\Phi_0$.  However, other parameters also contribute to the efficiency of the dynamo and it is possible to find supercritical solutions with $\Phi_0 = 1$.  One such parameter is the penetration depth of the BMRs, $r_s$ (see \S\ref{sec:STABLE}).  Here, deeper is better.  If we take the solution S1 and move the penetration depth up to 0.93$R$ instead of 0.9$R$ (Case S2 in Table 1), the dynamo dies (becomes subcritical).  

\begin{table}
\caption{Simulation Summary.  The S series of simulations uses the sawtooth pdf of eq.\ (\ref{eq:sawtooth}) while the L series uses the lognormal pdf of eq.\ (\ref{eq:lognormal}). Root-mean-square (rms) values listed in columns 8-10 are based on integrals over the entire computational volume and are quoted for the mean toroidal field ($B_{tor}$), the mean poloidal field ($B_{pol}$), and the non-axisymmetric field component ($B_{nax}$), which is mainly composed of BMRs.}

\begin{center}
\begin{tabular}{lccccccccc}
\hline
Case & $\Phi_0$ & $r_s$ & $s_a$ & $\eta_{top}$ & $\tau_p$ & $\tau_s$ & $B_{tor}$ & $B_{pol}$ & $B_{nax}$ \\
     &          &       &       & (cm$^2$ s$^{-1}$) & (days) & (days) & (rms) & (rms) & (rms) \\
\hline
S1 & 5 & 0.9 & 1.5 & $3 \times 10^{12}$ & -- & -- & 22 kG & 250 G & 720 G \\
S2 & 5 & 0.93 & 1.5 & $3 \times 10^{12}$ & -- & -- & subcritical & -- & -- \\ 
S3 & 5 & 0.93 & 2.5 & $3 \times 10^{12}$ & -- & -- & subcritical & -- & -- \\ 
S4 & 2 & 0.9 & 2.5 & $3 \times 10^{12}$ & -- & -- & 6.8 kG & 79 G & 200 G \\
S5 & 2 & 0.9 & 1.5 & $3 \times 10^{12}$ & -- & -- & subcritical & -- & -- \\ 
S6 & 1 & 0.9 & 1.5 & $3 \times 10^{12}$ & -- & -- & subcritical & -- & -- \\ 
S7 & 1 & 0.9 & 2.5 & $3 \times 10^{12}$ & -- & -- & subcritical & -- & -- \\ 
S8 & 1 & 0.9 & 2.5 & $10^{12}$ & -- & -- & subcritical & -- & -- \\ 
\hline
L1 & 1 & 0.9 & 2.5 & $3 \times 10^{12}$ & 2 & 3 & subcritical & -- & -- \\ 
L2 & 1 & 0.9 & 2.5 & $10^{12}$ & 2 & 3 & 11 kG & 130 G & 300 G \\
L3 & 1 & 0.85 & 4 & $10^{12}$ & 1 & 1.5 & 72 kG & 840 G & 960 G \\
\hline
\end{tabular}
\end{center}

\label{table1}
\end{table}

Another parameter that affects the efficiency of the dynamo is the spacing between spot pairs, $s_a$.  Recall from \S\ref{sec:STABLE} that $s_a$ is a nondimensional number that gives the distance between the two opposite polarity components of a BMR relative to the radius of the individual spots.  Thus, a large value of $s_a$ implies widely spaced spot pairs, which is beneficial for the dynamo because it maximizes the axisymmetric component of the poloidal flux and it minimizes local cancellation, allowing more trailing flux to reach the poles.  A comparison of cases S4 and S5 in Table 1 demonstrates that increasing $s_a$ from 1.5 to 2.5 can make the difference between a subcritical and a supercritical dynamo.  However, this is not always the case; compare also cases S2 and S3 and cases S6 and S7.

Since the dynamo must overcome diffusion to achieve supercriticality, a reduction in the diffusion can also be beneficial, particularly in the upper CZ where $\eta$ is largest (Fig.\ \ref{fig:ingredients}$c$).  This was not enough to revive Case S7; Case S8 is also subcritical even though the value of $\eta_{top}$ is decreased by a factor of three.  However, when combined with more frequent BMR emergence, which we achieved with the lognormal pdf (Fig.\ \ref{fig:spotmaker}$c$), lower $\eta_{top}$ did yield supercritical solutions, even for $\Phi_0 = 1$; see Cases L2 and L3 in Table 1.  The benefit of lower diffusion is demonstrated unambiguously by comparing cases L1 and L2, which both use the same lognormal emergence pdf.  The benefit of more frequent BMR emergence is demonstrated unambiguously by comparing cases S8 and L2.

\begin{figure}
\begin{center}
\includegraphics*[width=\linewidth]{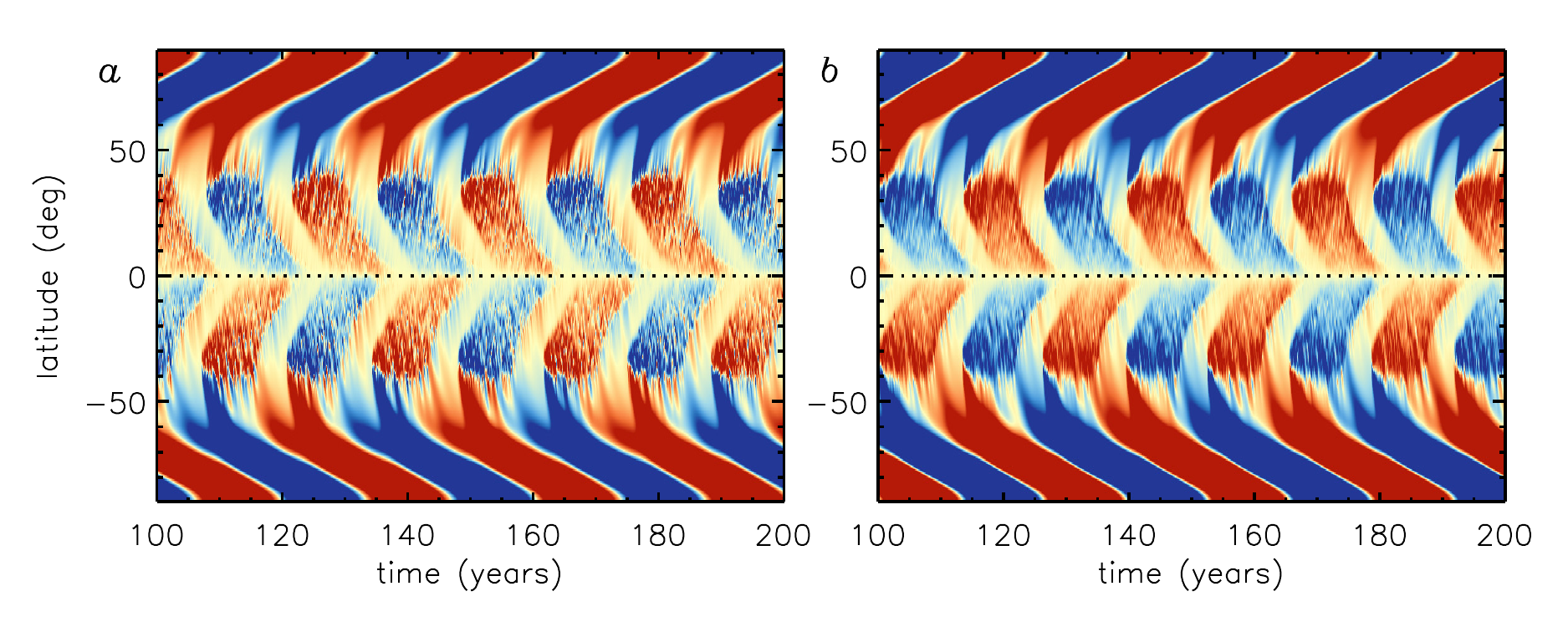}
\end{center}
\caption{As in Fig.\ \ref{fig:bfly}$a$ but for Cases ($a$) L2 and ($b$) L3, spanning a century of simulated time.  Saturation values for the color table are $\pm$ 50 G and $\pm$ 200 G respectively.
\label{fig:bflies}}
\end{figure}

The difference in dynamo efficiency influences not only the growth or decay rate of the dynamo, but also the nonlinear saturation level.  This is demonstrated most dramatically by Case L3 (Table 1).  This case has all the features that were shown above to be beneficial, including deep penetration of BMRs, wide spacing of BMR polarity components, low diffusion, and frequent BMR emergence, with a mean interval between spot appearances of $\tau_s = 1$ day and a mode of $\tau_p = 1.5$ days (Table 1).  Even though it has the same quenching field strength as all the other cases, $B_q = 10^5$ G, and a relatively low flux amplification factor of $\Phi_0 = 1$, it achieves stronger fields (more magnetic energy) than all of the other cases.  This includes case S1 (discussed in \S\ref{sec:flagship}), which has $\Phi_0 = 5$.  Furthermore, though Cases L2 and L3 have the same value of $\Phi_0$, the latter has much stronger fields.  It also has a higher ratio of $B_{pol}/B_{nax}$, approaching unity.  This can be attributed to the large value of $s_a$, which maximizes the mean poloidal field associated with each BMR because the two polarities have minimal overlap in latitude.

\begin{figure}
\begin{center}
\includegraphics*[width=\linewidth]{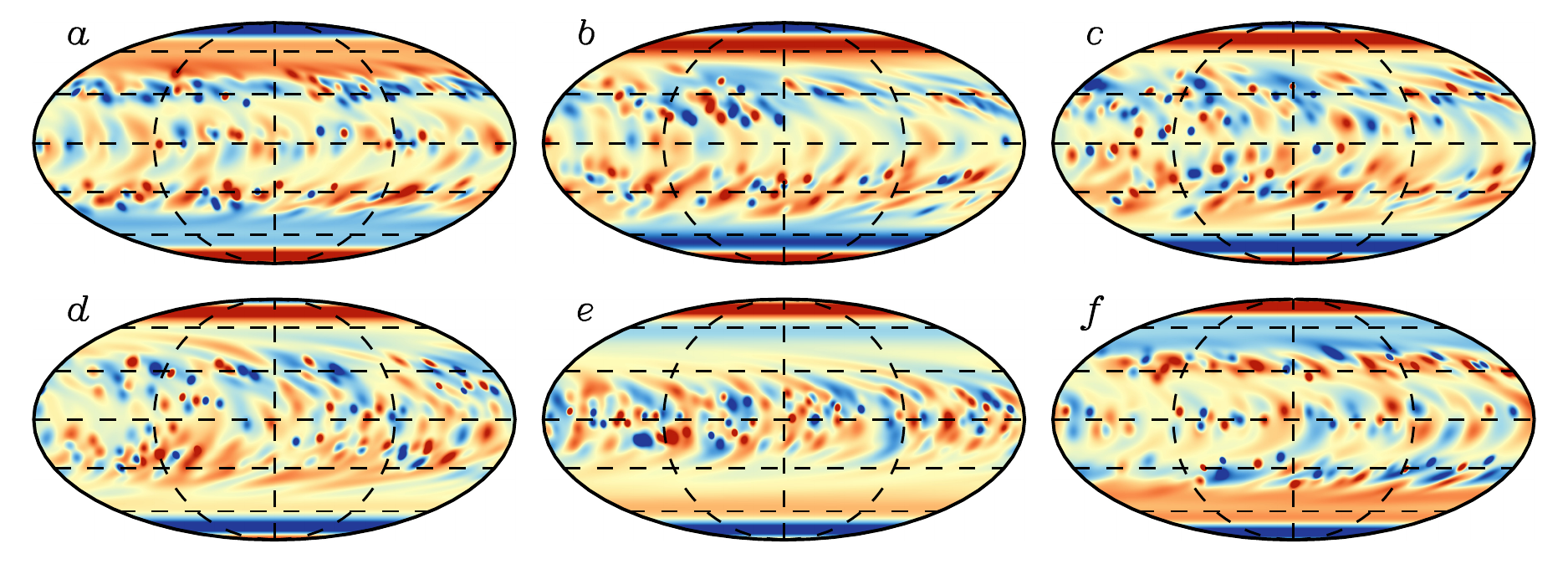}
\end{center}
\caption{As in Fig.\ \ref{fig:ss} but for Case L3. The saturation level for the color table is $\pm$ 1kG. \label{fig:ss2}}
\end{figure}

Fig.\ \ref{fig:metoo}$b$ shows ME$_{nax}$ in cases S8, L1, L2, and L3 for comparison.  All were started from the same initial conditions, obtained from the supercritical Case S4 (Table 1).  Note that Case L2 in particular has more magnetic energy than its progenitor, S4 (see also Table 1).  This demonstrates that lowering the diffusion by a factor of three more than makes up for lowering $\Phi_0$ by a factor of two.  Though the two supercritical cases in this plot, L2 and L3, look somewhat irregular at early times, they both settle into a steady magnetic cycle similar to that described in \S\ref{sec:flagship} for Case S1.  This is demonstrated in Fig.\ \ref{fig:bflies}.  The most apparent difference between these two cases (Figs.\ \ref{fig:bflies}$a$,$b$) is that case L3 ($b$) has a smoother distribution of low-latitude poloidal flux.  This can be attributed to the more frequent BMR emergence rate, which yields more flux patches at any given time.  Furthermore, the lower diffusion in both Cases L2 and L3 relative to Case S1 leads to stronger, more compact fields at low latitudes that more closely track the emergence sites; compare Fig.\ \ref{fig:bflies} with Figs.\ \ref{fig:scalar}$b$ and \ref{fig:bfly}$a$.  

The stronger, more compact fields at low latitudes in Case L3 relative to Case S1 is apparent in Fig.\ \ref{fig:ss2}.  Compare this to Fig.\ \ref{fig:ss}.  Each individual spot of a BMR takes longer to disperse so at any given time, there are more spots that are strong and localized.  The flux distribution in Case L2 (not shown) is similar.  Yet, this change in the flux distribution has little effect on the overall characteristics of the magnetic cycles.  The cycle period in all three cases is similar: 13.1 years for S1 and L3, and 13.6 years in L2.  The other supercritical solution, Case S4, has a cycle period of 13.0 years.

Though a detailed comparison with observations lies outside the scope of this paper, we note that a cursory comparison of our simulations with solar magnetograms and SFT models suggests that cases L2 and L3 produce more realistic surface flux distributions.   Their lower diffusion and higher frequency of emergence events tends to produce strong flux patches with a wider range of areas that extends below 5 square degrees, reminiscent of observed magnetograms \citep{upton14a}.

As discussed in \S\ref{sec:fluxdep}, the SpotMaker algorithm effectively assumes that only a small portion of a progenitor toroidal loop emerges through the surface so the depletion of mean flux is minimal.  In reality, a significant portion of the mean flux in the lower CZ/tachocline would be redistributed throughout the CZ and outside of the domain.  Taking this tachocline flux depletion into account would likely inhibit the dynamo; it is unclear whether or not supercritical solutions could still be found for $\Phi_0=1$.  It would also reduce the ratio of mean toroidal to poloidal field.  These issues are indeed important to investigate but they are subtle, and beyond the scope of this paper.

\section{Summary}
\label{sec:summary}

We have described a novel 3D Babcock-Leighton/Flux Transport solar dynamo model and we have presented some initial results for axisymmetric, kinematic flow fields.  These initial results provide a baseline for subsequent studies that will incorporate 3D flow fields and Lorentz-force feedbacks.   The name of the new model is STABLE, reflecting its close ties to both surface flux transport models (ST) and (A) Babcock-Leighton (BLE) dynamo models.

The use of kinematic, axisymmetric flow fields implies that the evolution of mean fields reduces to an equivalent 2D FTD model, though there is no explicit $\alpha$-effect (\S\ref{sec:STABLE}).  Instead, poloidal field is generated by the BL mechanism in response to the spontaneous appearance of BMRs, as in the 2D FTD model of \cite{munoz10}.  Our model is also similar to the 3D FTD model of \cite{yeate13}, although they use a more sophisticated flux emergence algorithm based on imposed rising, helical, flow fields.  We plan to implement a similar algorithm in the future.  Until then, the current BMR deposition approach may be regarded as the limit in which emergent toroidal flux structures decouple quickly from their roots in the deep CZ.  In order to verify the STABLE model (\S\ref{sec:verification}), we provisionally replaced this BMR deposition algorithm with an axisymmetric BL source term and we were able to reproduce the 2D FTD benchmark CS$^\prime$ defined by \cite{jouve08}.

The STABLE model exhibits many promising features that are in good agreement with solar observations.  Like other FTD models, it sustains regular magnetic cycles with equatorward propagation of toroidal flux at low to mid latitudes near the base of the CZ.  However, unlike many other mean-field and convective dynamo models, there is no need to use this subsurface toroidal flux as a proxy for sunspot number.  Instead, sunspots/BMRs appear at the surface of the STABLE model explicitly and, in the special case of axisymmetric flow fields, can be tracked by means of the non-axisymmetric component of the magnetic energy (Fig.\ \ref{fig:scalar}).  The evolution of the radial magnetic field at the surface is similar to SFT models, with the distortion and dispersal of low-latitude, tilted BMRs by DR, MC, and turbulent diffusion producing high-latitude, axisymmetric bands of poloidal flux that migrate poleward and eventually reverse the polar fields.  This process is essential for the operation of the dynamo.

As in other FTD models, the period and amplitude of the magnetic cycles are largely determined by the meridional flow speed and the quenching of the poloidal source (through the $B_q$ factor in eq.\ (\ref{eq:flux})).  However, other parameters do play a significant role.  For example, though the MC and $B_q$ are the same in all simulations presented here (with the exception of the benchmark in \S\ref{sec:verification}), the total magnetic energy produced by the dynamo can vary by more than two orders of magnitude (see Table 1).  The cycle period is less sensitive to the BMR structure, emergence rate, and diffusion, but it can still vary significantly, from 13.0 years in Case S4 to 13.6 years in Case L2.

Factors that influence the dynamo efficiency include the emergence rate and penetration depth of BMRs, and the spacing between the two polarity components of a BMR.  Poloidal field generation is more efficient if each BMR is wider and deeper, and if the number of BMRs is increased by increasing the frequency of emergence (see Fig.\ \ref{fig:spotmaker}$c$).  Lower diffusion also enhances the dynamo efficiency and makes the distribution of radial flux at the surface more intermittent (Fig.\ \ref{fig:ss2}).  Any of these factors can make the difference between a subcritical and supercritical solution (\S\ref{sec:supercritical}).

One deficiency of the model in its current state is the long time it takes for residual flux from mid-latitude BMRs to migrate poleward and reverse the polar fields (Figs.\ \ref{fig:bfly} and \ref{fig:bflies}).  This does not agree well with solar observations, both in terms of the migration speed and the phasing between low-latitude toroidal flux as traced by sunspots and high-latitude polar field reversals.  As discussed in \S\ref{sec:overview}, this is likely due to a relatively low speed for the high-latitude meridional flow and will be addressed in future versions of STABLE.  Other features of the model can also be calibrated to improve agreement with solar observations, including the time lag pdf of Fig.\ \ref{fig:spotmaker}$c$, which may depend on the phase of the cycle.

Boundary conditions appear to be important for the solar dynamo.  In particular, it is clear from solar observations that substantial magnetic flux passes through the surface of the sun and that the shearing of this poloidal flux by the surface DR generates toroidal flux in each hemisphere by means of the $\Omega$-effect.  CS15 argue that this is the dominant source of mean toroidal flux in the Sun and can account for all the flux that emerges in BMRs.  We find that this is indeed the case for our FTD models.  However, we find that the role of turbulent diffusion is somewhat different than that envisioned by CS15 (\S\ref{sec:cs}).  They modeled turbulent diffusion as an effective drag term that inhibited the generation of net toroidal flux in each hemisphere $\Phi_{NH,SH}(t)$.  We find instead that it promotes the generation of net flux by selectively dissipating residual oppositely-signed flux from the previous cycle.  Thus, instead of inducing a negative phase shift in $\Phi_{NH,SH}(t)$ as argued by CS15, we find that diffusion induces a positive phase shift, causing the maxima and reversals in $\Phi_{NH,SH}(t)$ to occur later than they would without diffusion.

Future model developments will include flux transport and amplification from 3D convective flow fields, Lorentz force feedbacks, and flows induced by enhanced radiative cooling in active regions, which is expected to play a significant role in dynamo saturation.  These developments may produce non-asymmetric dynamo modes, magneto-shear instabilities, and torsional oscillations that will be explored in future papers.  Furthermore, we intend to explore the SFT aspects of STABLE and its potential for forecasting future magnetic activity by assimilating data from photospheric magnetograms.  We believe this class of 3D BL/FTD models shows great promise as a ``Solar Dynamo Frontier''.

The authors wish to thank the NCAR Advanced Study Program for funding KT's visit to HAO/NCAR in support of this research.  We also thank Mausumi Dikpati, Gopal Hazra, Bidya Karak, and Lisa Upton for many enlightening conversations about this project.  The computations were performed using resources provided by NASA's High End Computing (HEC) program (Pleiades) and NCAR (Yellowstone).  The National Center for Atmospheric Research is sponsored by the National Science Foundation.




\end{document}